\documentclass[journal=jctcce,manuscript=article]{achemso}
\setkeys{acs}{articletitle = true}

\usepackage[version=3]{mhchem} % Formula subscripts using \ce{}
\usepackage[dvipsnames]{xcolor}
\usepackage{adjustbox}
\usepackage{tikz, pgfplots, pgfplotstable}
\pgfplotsset{compat=1.8}
\usepackage{pdfpages}
\usetikzlibrary{arrows.meta,external}
\tikzset{%
  >={Latex[width=2mm,length=2mm]},
            base/.style = {draw=black,
                           minimum width=2cm, 
                           minimum height=1cm, 
                           text centered}
}
\usepackage[utf8]{inputenc}  
\usepackage[T1]{fontenc}
\usepackage{gensymb}
\usepackage{changes}

\usepackage{lipsum}

\definecolor{cl_pbe}{HTML}{000000}
\definecolor{cl_beef}{HTML}{9A6324}
\definecolor{cl_tpss}{HTML}{FFE119}
\definecolor{cl_scan}{HTML}{3CB44B}
\definecolor{cl_pbe0}{HTML}{4363D8}
\definecolor{cl_hse06}{HTML}{F58231}
\definecolor{cl_b3lyp}{HTML}{E6194B}

\newcommand{\tcm}{$\eta_\text{\,TCM}$}

\author{Bj{\"o}rn Kirchhoff}
\affiliation[HI]
{Science Institute and Faculty of Physical Sciences, University of Iceland, VR-III, Hjarðarhagi 2, 107 Reykjav\'{\i}k, Iceland}
\alsoaffiliation[UUlm]
{Institute of Electrochemistry, Ulm University, Albert-Einstein-Allee 47, 89081 Ulm, Germany}
\author{Aleksei Ivanov}
\affiliation[HI]
{Science Institute and Faculty of Physical Sciences, University of Iceland, VR-III, Hjarðarhagi 2, 107 Reykjav\'{\i}k, Iceland}
\author{Egill Sk{\'u}lason}
\affiliation[HI]
{Science Institute and Faculty of Industrial Engineering, Mechanical Engineering and Computer Science, University of Iceland, Hjarðarhagi 2, 107 Reykjavík, Iceland}
\author{Timo Jacob}
\affiliation[UUlm]
{Institute of Electrochemistry, Ulm University, Albert-Einstein-Allee 47, 89081 Ulm, Germany}
\alsoaffiliation[HIU]
{Helmholtz-Institute Ulm (HIU) Electrochemical Energy Storage, Helmholtz-Straße 16, 89081 Ulm, Germany}
\alsoaffiliation[KIT]
{Karlsruhe Institute of Technology (KIT), P.O. Box 3640, 76021 Karlsruhe, Germany}
\author{Donato Fantauzzi}
\affiliation[HI]
{Science Institute and Faculty of Physical Sciences, University of Iceland, VR-III, Hjarðarhagi 2, 107 Reykjav\'{\i}k, Iceland}
\author{Hannes J{\'o}nsson}
\email{hj@hi.is}
\affiliation[HI]
{Science Institute and Faculty of Physical Sciences, University of Iceland, VR-III, Hjarðarhagi 2, 107 Reykjav\'{\i}k, Iceland}

\title{Assessment of the Accuracy of Density Functionals for Calculating Oxygen Reduction Reaction on Nitrogen Doped Graphene}

\abbreviations{DFT}

\begin{document}

\begin{tocentry}

   \includegraphics[width = 0.49\columnwidth]{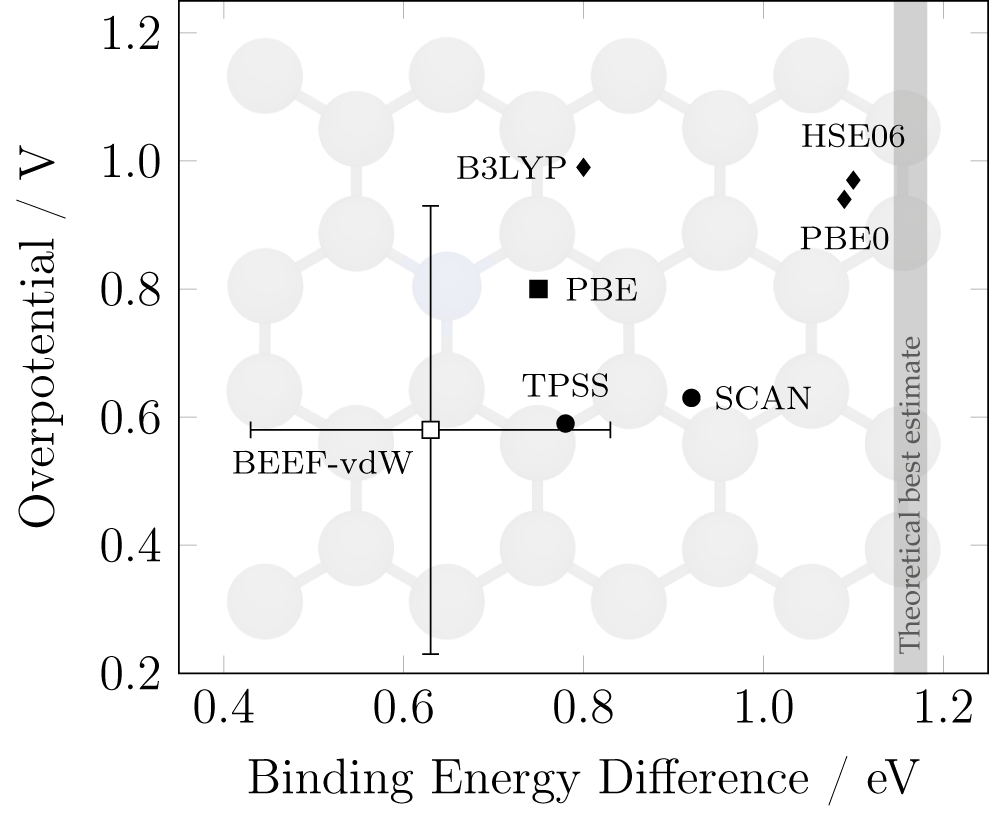}

Calculated overpotential for ORR on 3.1\% N doped graphene using various energy functionals as a function of the calculated energy difference of oxygen adatom at on-top site and
bridge site. The best estimate of the binding energy difference obtained from quantum Monte Carlo simulations is indicated with a gray area.
A standard deviation error bar obtained for both quantities using the BEEF-vdW functional are also shown.

\end{tocentry}

\begin{abstract}
%The accuracy of several density functionals and Perdew-Zunger Self-Interaction Correction (SIC) is benchmarked for calculations of the adsorption energy of oxygen on graphene against Diffusion Monte Carlo values. The PBE0 and HSE06 hybrid functionals as well as SIC are found to reproduce the benchmark results well. PBE, BEEF-vdW, TPSS, SCAN, and B3LYP produce errors of \textit{ca.}~25--40~\%. The thermochemical overpotential \tcm\ of the oxygen reduction reaction on a periodic nitrogen-doped graphene model calculated with PBE0, HSE06, or B3LYP is \textit{ca.}~1~V, while the other functionals underestimate \tcm\ by up to 0.4~V compared to HSE06. Solvation and an increase of the dopant concentration are found to decrease \tcm\ calculated with HSE06. GGA and meta-GGA functionals are found to overestimate the binding energy of the *O intermediate, which can lead to different overall trends in the calculated overpotential as compared to hybrid functionals. This overbinding issue can thus lead to different overall trends when using GGA functionals or hybrids. Lastly, molecular flake models are shown to behave erratically due to finite size effects and geometric distortions during relaxation.
%
Experimental studies of the oxygen reduction reaction (ORR) at nitrogen doped graphene electrodes have reported a remarkably low overpotential, on the order of 0.5 V, similar to Pt based electrodes. Theoretical calculations using density functional theory have lent support for this claim. However, other measurements have indicated that transition metal impurities are actually responsible for the ORR activity, thereby raising questions about the reliability of both the experiments and the calculations. In order to assess the accuracy of the theoretical calculations, various generalized gradient approximation (GGA), meta-GGA and hybrid functionals are employed here and calibrated against high-level wave function based coupled cluster calculations (CCSD(T)) of the overpotential as well as self-interaction corrected density functional calculations and published quantum Monte Carlo calculations of O adatom binding to graphene. The PBE0 and HSE06 hybrid functionals are found to give more accurate results than the GGA and meta-GGA functionals, as would be expected, and for low dopant concentration, 3.1\%, the overpotential is calculated to be 1.0~V. The GGA and meta-GGA functionals give a lower estimate by as much as 0.4~V. When the dopant concentration is doubled, the overpotential calculated with hybrid functionals drops, while it increases in GGA functional calculations. The opposite trends result from different potential determining steps, the *OOH species being of central importance in the hybrid functional calculations while the reduction of *O determines the overpotential obtained in GGA and meta-GGA calculations. The results presented here are mainly based on calculations of periodic representations of the system, but a comparison is also made with molecular flake models which are found to give erratic results due to finite size effects and geometric distortions during energy minimization. The presence of the electrolyte has not been taken into account explicitly in the calculations presented here, but is estimated to be important for definitive calculations of the overpotential.
%obtained from the hybrid functional calculations by 0.2~V.
%
%   add ferromagnetic ordering when 6.2% doping when HSE06 used, but not PBE
%    HSE06 still shows *OOH reduction to be potential determining step, while for GGA functionals it is the *O reduction step. 
\end{abstract}


\section{Introduction}

Doped graphene has attracted attention over the past decade as a metal-free catalyst\cite{zhang2015,hu2016,sturala2018}
and is being explored as a possible replacement for expensive and rare platinum group metals for the oxygen reduction reaction (ORR) in fuel cells\cite{ge2015}. 
%In particular, these materials are explored as replacement for expensive and rare platinum group metals as oxygen reduction reaction (ORR) electrocatalysts in fuel cells\cite{ge2015}. 
Qu {\it et al.} synthesized a few-layer nitrogen-doped graphene (NG) by chemical vapor deposition on a \ce{SiO2}/\ce{Ni} substrate and used
it as the cathode in ORR experiments.
They reported an onset potential similar to commercial platinum catalysts under alkaline conditions 
%while showing higher electrocatalytic activity, 
and better resistance to methanol and CO poisoning, as well as good cycling stability.\cite{qu2010} 
Since then, several experimental studies have reported on high ORR activity in NG as has been reviewed 
by Wang \textit{et al.}\cite{wang2012} and Zhang \textit{et al.}\cite{zhang2015} 
Generally, NG is reported to catalyze the \ce{4e-} ORR mechanism at similar or slightly lower catalytic activity than Pt under alkaline conditions while having higher durability. 
%Note that the influence of transition metal impurities on the catalytic activity of NG is subject to ongoing debate\cite{ambrosi2012,wang2013,masa2013,masa2015,wang2020} but is outside the scope of this work.

The possible influence of transition metal impurities on the catalytic activity measured for NG is, however, a subject of ongoing debate. Some widely used preparation methods\cite{hummers1958,marcano2010} are based on a permanganate oxidant and the Mn atoms can be present
in the product even after washing\cite{wang2013}. 
%Ambrosi and co-workers have shown that removal of transition metal contaminations post-synthesis is impossible using wet-chemistry methods and that 
Extreme conditions 
%(1000 \degree C in chlorine atmosphere) are 
have in some cases been used to reduce the amount of metal residue and this has been found to result in reduced ORR activity \cite{ambrosi2012}. 
NG synthesized under explicitly metal-free conditions has been reported to give higher ORR overpotential.\cite{masa2013}

Theoretical studies of ORR on NG have been conducted using various approaches.
Classical dynamics simulations with energy and atomic forces estimated using electron density functional theory (DFT) have been used to 
study the relative stability of the various intermediates in ORR as well as free energy barriers for the reaction steps.\cite{ikeda2008,okamoto2009,yu2011}  
The first theoretical estimate of the overpotential was presented by Studt\cite{studt2013} using the thermochemical model (TCM)\cite{norskov2004}.
There, the overpotential, \tcm\ , is estimated from the free energy of the intermediates and the computational hydrogen electrode.
Using the BEFF-vdW functional,\cite{wellendorff2012} which is of the generalized gradient approximation (GGA) functional form,
Studt obtained a value of \tcm\ = 0.72 V for NG with a dopant concentration of 6.2\% 
and identified the potential determining step to involve reduction of *O adatom,
but pointed out that solvation effects need to be taken into account in order to obtain a more reliable estimate.\cite{studt2013}
We note that the TCM also does not include kinetic effects, \textit{i.e.} the free energy barriers of the elementary reaction steps.\cite{skulason2017} 
Reda {\it et al.} later carried out an extensive study using the same functional and method to estimate  \tcm\ for a range of dopant concentrations and to assess the effect of solvation by including ice-like layers of water molecules.\cite{reda2018}. 
% they found ...
Using another functional of the GGA form, the PBE functional,\cite{perdew1996} 
Li \textit{et al.} reported a lower value for the onset potential, 0.45~V, from an extrapolation of calculated results for N-doped nanoribbon models\cite{li2014}. 
G\'{\i}slason and Sk\'ulason\cite{margislason_catalytic_2019} obtained a \tcm\ a value of 0.57~V for the overpotential using the RPBE functional,\cite{hammer1999}
a version of the PBE functional adjusted to give adsorption energy in better agreement with experimental measurements.
Other calculations using the PBE functional have reported even lower values for the overpotential, such as 0.48~V for a model containing a cluster of three N dopants.\cite{sinthika2018}
%Their study did not take into account solvation. Chai \textit{et al.} report a \tcm\ of 0.43~V for a strained NG containing a Stone-Wales defect using the HCTH functional\cite{chai2014}. Sinthika and co-workers proposed a descriptor approach for sp$^2$ hybridized materials based on occupations of p$_\text{z}$ orbitals\cite{sinthika2018}. Using the PBE functional, they extrapolate a minimum \tcm\ of 0.48~V for an NG infinite-sheet model containing a cluster of three N dopants.

The above studies were carried out using periodic, infinite-sheet or ribbon models of the system, 
but some other studies have used calculations of 
finite, molecular flake models to estimate the overpotential of ORR on NG. These studies have invariably been carried out with hybrid
density functionals where some fraction of exact exchange is included in the functional form. 
These functionals are generally considered to give more accurate results than
GGA functionals. Using the B3LYP hybrid functional,\cite{stephens1994,kim1994} Zhang and Xia 
%is often credited with providing theoretical proof that the \ce{4 e-} mechanism should be dominant for NG\cite{zhang2011}. Zhang \textit{et al.} used the B3LYP functional as well to study the ORR on a 
calculated the overpotential for a
flake model with a Stone-Wales defect\cite{zhang2011}, and in a combined experimental and theoretical study, Jiao \textit{et al.} reported \tcm\ values for graphene doped with various types of atoms (N, B, P, S, O).\cite{jiao2014} 
They found that B-doped graphene exhibits similarly good ORR performance as NG with \tcm\ values similar to a commercial Pt catalyst.

%The methods and graphene model systems used in the various reported computational studies of this topic are varied. Generally, two main approaches have been used. One group of studies is based on periodic calculations on either infinite-sheet or nanoribbon models using the PBE\cite{perdew1996} functional within the generalized gradient approximation (GGA)\cite{okamoto2009,kim2011,yu2011,boukhvalov2012,chai2014,choi2014,chai2017,sinthika2018}. The other group of studies relied on capped molecular models in combination with the B3LYP\cite{stephens1994,kim1994} hybrid functional\cite{sidik2006,zhang2011,zhang2012,jiao2014}. Some noteworthy outliers are the studies by Studt and Reda, who performed periodic dispersion-corrected BEEF-vdW\cite{wellendorff2012} calculations\cite{studt2013,reda2018},  Gíslason and Skúlason who used RPBE and an infinite-sheet model\cite{margislason_catalytic_2019}, Vazquez-Arenas \textit{et al.} who used PBE0\cite{perdew1996_2,adamo1999} on a molecular model\cite{vazquez2016}, Xie \textit{et al.} who used the PBE-adjacent PW91\cite{burke1998} in combination with an infinite-sheet model\cite{xie2018}, as well as Ikeda\cite{ikeda2008} and Chai\cite{chai2014} and their respective co-workers who employed the HTCH functional\cite{hamprecht1998} in periodic DFT calculations.

The theoretical results discussed briefly above essentially fall into two classes, namely calculations of extended, periodic models of the system
using energy functionals of the GGA form, or hybrid energy functional calculations of finite, molecular-like models. 
Apparently, both approaches give similarly good results indicating that NG can have ORR overpotential close to that of Pt catalysts.
This could be interpreted as an indication that
GGA and hybrid functionals are equally applicable to these studies and that periodic and finite models of the system give equivalent results.
This is suprising since benchmark diffusion Monte Carlo simulations of the binding of an O adatom on a periodic model of graphene have shown
significant errors in the binding energy obtained from GGA functionals.\cite{hsing2012}
Janesko {\it et al.} expanded on this benchmark study\cite{janesko2013} by testing also meta-GGA and hybrid functionals, showing that hybrid 
functionals are significantly more accurate for this application, especially the PBE0\cite{perdew1996_2,adamo1999} and HSE06\cite{krukau2006} functionals.
%\added{In another recent study, Mahler \textit{et al.} presented results on surface reactions where hybrids fail to improve on GGA functionals which highlights the necessity to re-evaluate the performance of the various DFT rungs for any new material class.\cite{mahler_why_2017}}
In another recent study, Mahler \textit{et al.} presented results on surface reactions where hybrids fail to improve on GGA functionals which highlights the necessity to re-evaluate the performance of the various DFT rungs for any new material class.\cite{mahler_why_2017}
The question, therefore, arises how the overpotential for ORR predicted by hybrid functionals for periodic model systems of NG compare with
the values previously obtained using GGA functionals, and whether the finite, flake models of NG give similar results as the extended, 
periodic models. 

%To summarize, computational studies on NG have come to vastly different conclusions regarding the ORR activity of the studied model systems. Studies rely on generalized gradient approximation (GGA) functionals (PBE in particular) if periodic model systems are used and on the B3LYP hybrid functional if molecular flake models are used. Note however that benchmarks recommend against these functionals. In a Diffusion Monte Carlo (DMC) study, Hsing and co-workers calculate adsorption of oxygen, fluorine, and hydrogen on a periodic graphene supercell and compared the results to LDA- and several GGA-DFT methods\cite{hsing2012}. They concluded that while the DFT calculations are able to reproduce H adsorption to a reasonable accuracy, O and F adsorption are poorly described using LDA and GGA functionals. In their study, O adsorption at the preferred bridge site is overestimated by \textit{ca.} 35~\% with PBE. Janesko and co-workers expand on this study by comparing Hsing's DMC results to selected meta-GGA and hybrid functionals as well as to the GGA functionals used in the original study\cite{janesko2013}. The group finds an analogous trend for the GGA functionals but highlight that results improve notably when hybrid functionals are used. The B3LYP functional is reported to produce worse results than other hybrids such as PBE0 and HSE06.

This article reports on a comprehensive study of various DFT functionals for calculating the overpotential for ORR on NG using both periodic as well as finite models. First, benchmark calculations of oxygen adsorption on undoped graphene are performed with a set of DFT functionals as well as explicitly self-interaction corrected GGA functional and the results compared with the diffusion Monte Carlo results.
Secondly, \tcm\ values for ORR on NG are reported for the various DFT functionals and both periodic and finite model systems. 
A remarkably large range of values is obtained, indicating what level of theory is needed to obtain best estimates of the overpotential 
within the TCM approximation.
The effect of solvation remains the largest uncertainity, as discussed in the context of several different estimates.

% ----------------------------------------------------------

\section{Methodology}

The calculations of the periodic representation of the system are carried out using a plane wave basis set with an energy cutoff of 600~eV
to represent valence electrons and the projector-augmented wave (PAW) method \cite{blochl_projector_1994,kresse1999} used to account for the effect of inner electrons. The energy functionals used in this study include the GGA functionals PBE\cite{perdew1996}, BEEF-vdw\cite{wellendorff2012}, as well as the meta-GGA functionals TPSS\cite{furche2006}, SCAN\cite{sun2015}, and hybrid functionals 
PBE0\cite{perdew1996_2,adamo1999}, HSE06\cite{krukau2006}, and B3LYP\cite{stephens1994,kim1994}. 
For BEEF-vdW, the implementation by Klimeš \textit{et al.} was used\cite{klimes2010,klimes2011}. 
If not stated otherwise, the simulation cell includes a 32-atom representation of the graphene and is constructed using lattice parameters (see Table S1) obtained from a \ce{C4} graphite bulk cell (P6$_3$/mmc spacegroup) optimized using a converged $11 \times 11 \times 3$ $k$-point grid. Wave functions are self-consistently optimized until the energy in subsequent iterations changes by less than $10^{-6}$~eV. Atomic coordinates are optimized until forces drop below $10^{-2}$~eV~\AA$^{-1}$. Gaussian-type finite temperature smearing is used to speed up convergence. The smearing width is chosen so that the electronic entropy remains below 1~meV. Real-space evaluation of the projection operators is used to speed up calculations of larger systems, using a precision of $10^{-3}$~eV~atom$^{-1}$. This scheme is also used with the isolated molecules to ensure consistency. 
The periodic images are separated by 14~\AA\ of vacuum and a dipole correction is applied perpendicular to the slab.
Omnidirectional dipole correction is used in the calculations of isolated molecules.
The calculations are performed with the Vienna {\it ab initio} simulation package (VASP) version 5.4.4.\cite{kresse1993, kresse1994, kresse1996, kresse1996_2}

DFT calculations of the finite model systems, \textit{i.e}. the flakes, are carried out using the PBE and PBE0 density functionals. Furthermore, 
a high-level wave function based approach, the coupled-cluster singles doubles and perturbative triples (CCSD(T)) method, is used to test the
accuracy. 
The domain-based local pair natural orbital (DLPNO) approximation\cite{dlpno1,dlpno2,dlpno3,dlpno4,dlpno5} is used to reduce the computational effort of CCSD(T)
calculations and a complete basis set extrapolation scheme\cite{extrapolation1,extrapolation2,extrapolation3} with the cc-pVDZ and cc-pVTZ\cite{dunning_gaussian_1989} basis sets is used to reduce the basis set error. 
The def2-TZVP triple-$\zeta$ basis set is used\cite{weigend_balanced_2005} in the DFT calculations and the
RIJCOSX\cite{rijcosx} approximation applied to speed up the exact exchange part in hybrid DFT and DLPNO-CCSD(T) calculations. 
The def2/J\cite{weigend_hartreefock_2008} auxiliary basis set is used there. The cc-pVTZ/C\cite{weigend_efficient_2002} auxiliary basis set is additionally used with DLPNO-CCSD(T) calculations.
The calculations are performed using ORCA software version 4.0.1.\cite{neese_orca_2012, neese_orca_2018}

\begin{figure}[h!]
	\centering
		\includegraphics[width=0.6\linewidth]{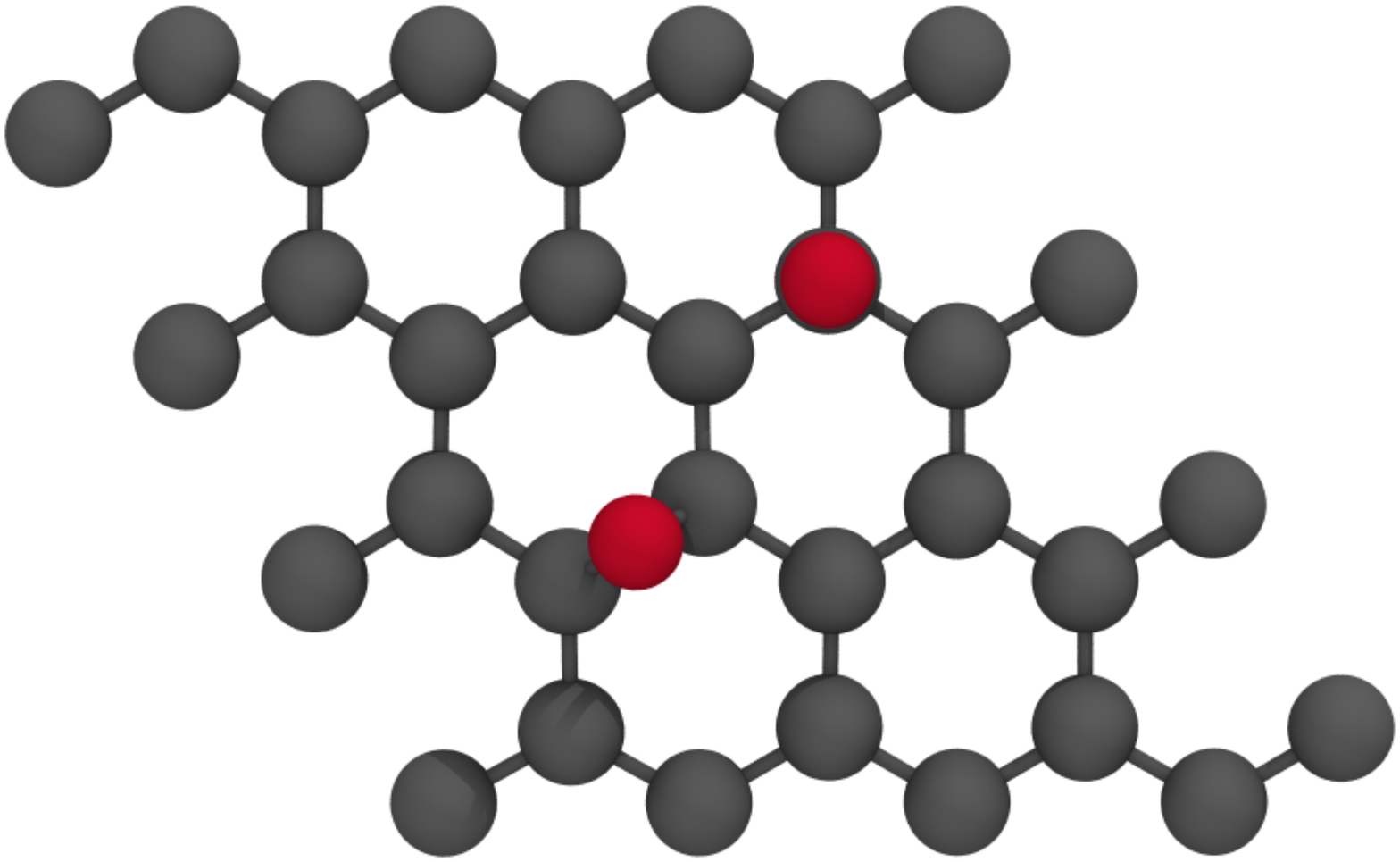}
	\caption{The 4x4 graphene simulation cell and location of the O-adatom at bridge and on-top sites.}
	\label{fgr:G_positions}
\end{figure}

The test calculations on oxygen adsorption on undoped graphene are performed using a rhombic $4 \times 4$ supercell with 32 atoms (G32),
as illustrated in figure \ref{fgr:G_positions}. 
Comparison is made with high-level diffusion Monte Carlo (DMC) results reported by Hsing \textit{et al.} with atomic coordinates provided by the authors. To ensure compatibility with the DMC calculations, calculations are carried out using the \textit{M}(0.5, 0) special $k$-point,
but converged \textit{\textbf{k}} grid calculations are also carried out with a $5\times5\times1$ $\Gamma$-centered grid. 
%The model system and the two calculated positions of the O adatom are illustrated in 
The lowest energy site for the oxygen adatom is the bridge position and the on-top site is significantly higher in energy.
Calculations focus on the energy difference between the two configurations.
% and corresponds to a first order saddle point according to frequency analysis.
%
% ----------------------------------   Figure 1 --------------------------------------------
%  --------------------------------------------------------------------
%

In order to gain further insight into the limitations of the accuracy of the GGA functionals,
an explicit self-interaction correction as proposed by Perdew and Zunger\cite{perdew1981}
 (PZ-SIC) is applied to the PBE functional in the calculations of the O-adatom configurations.
The calculations make use of a real-space grid representation of the valence electrons and PAW for inner electrons
as implemented in the GPAW software version 20.1.0\cite{enkovaara_electronic_2010}. 
A grid mesh of 0.15 Å is used and direct minimization over complex orbitals is carried out\cite{klupfel_importance_2011,lehtola2013,lehtola2016}. 
As has been established previously from atomization and band gap studies, the PZ-SIC is scaled by a half.\cite{klupfel_effect_2012}

\begin{figure}[h!]
	\centering
			\includegraphics[width=0.6\linewidth]{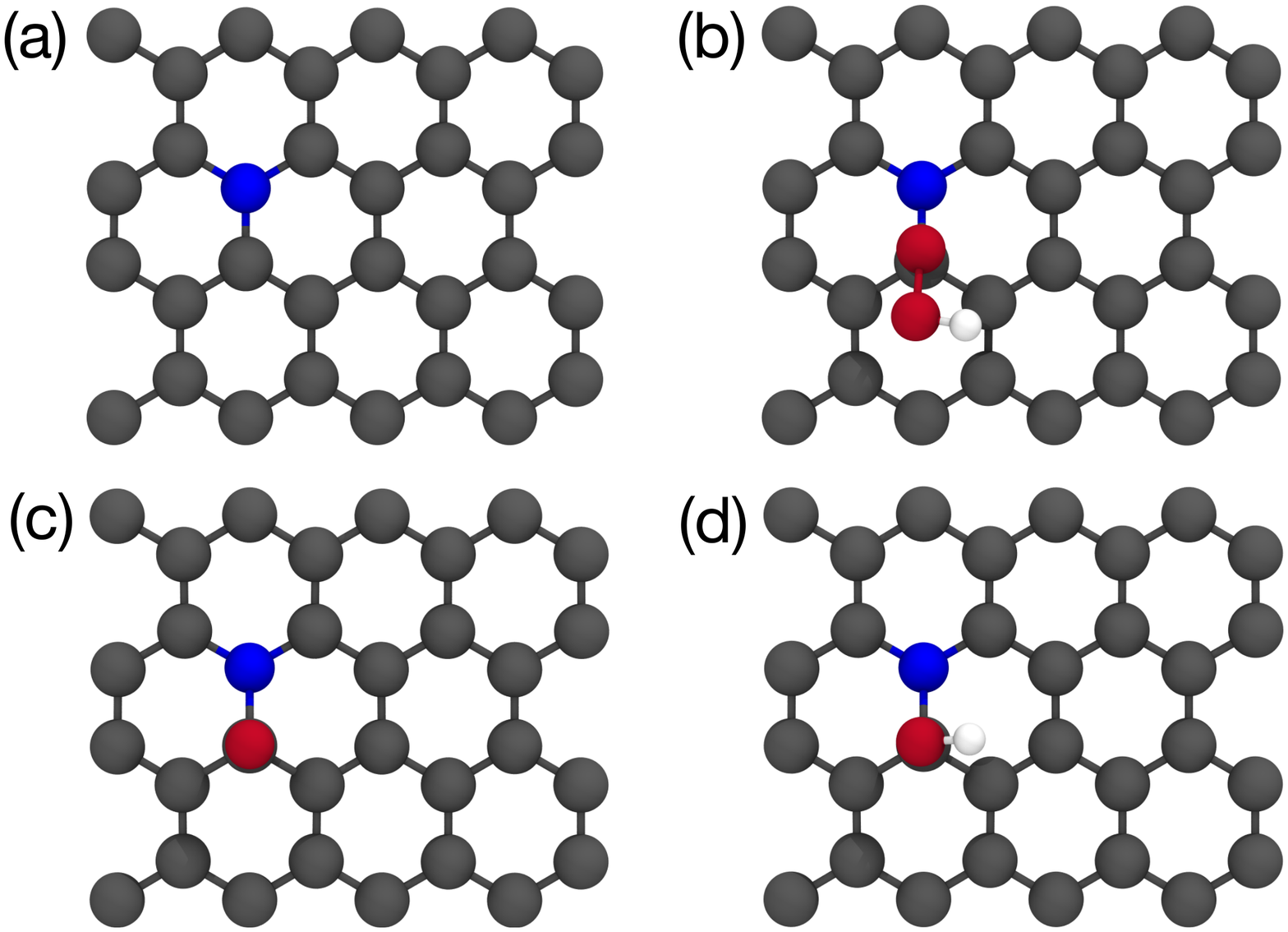}
	\caption{Simulation cell of the 3.1\% N-doped graphene (\textbf{a}) 
	and relaxed structures of the oxygen reduction reaction intermediates: *OOH (\textbf{b}), *O (\textbf{c}), and *OH (\textbf{d}).
	N atoms blue,  O atoms red,  H atoms white,  C atoms gray.
}	
	\label{fgr:dopedmodels}
\end{figure}

\begin{figure}[h!]
	\centering
	\includegraphics[width=0.6\linewidth]{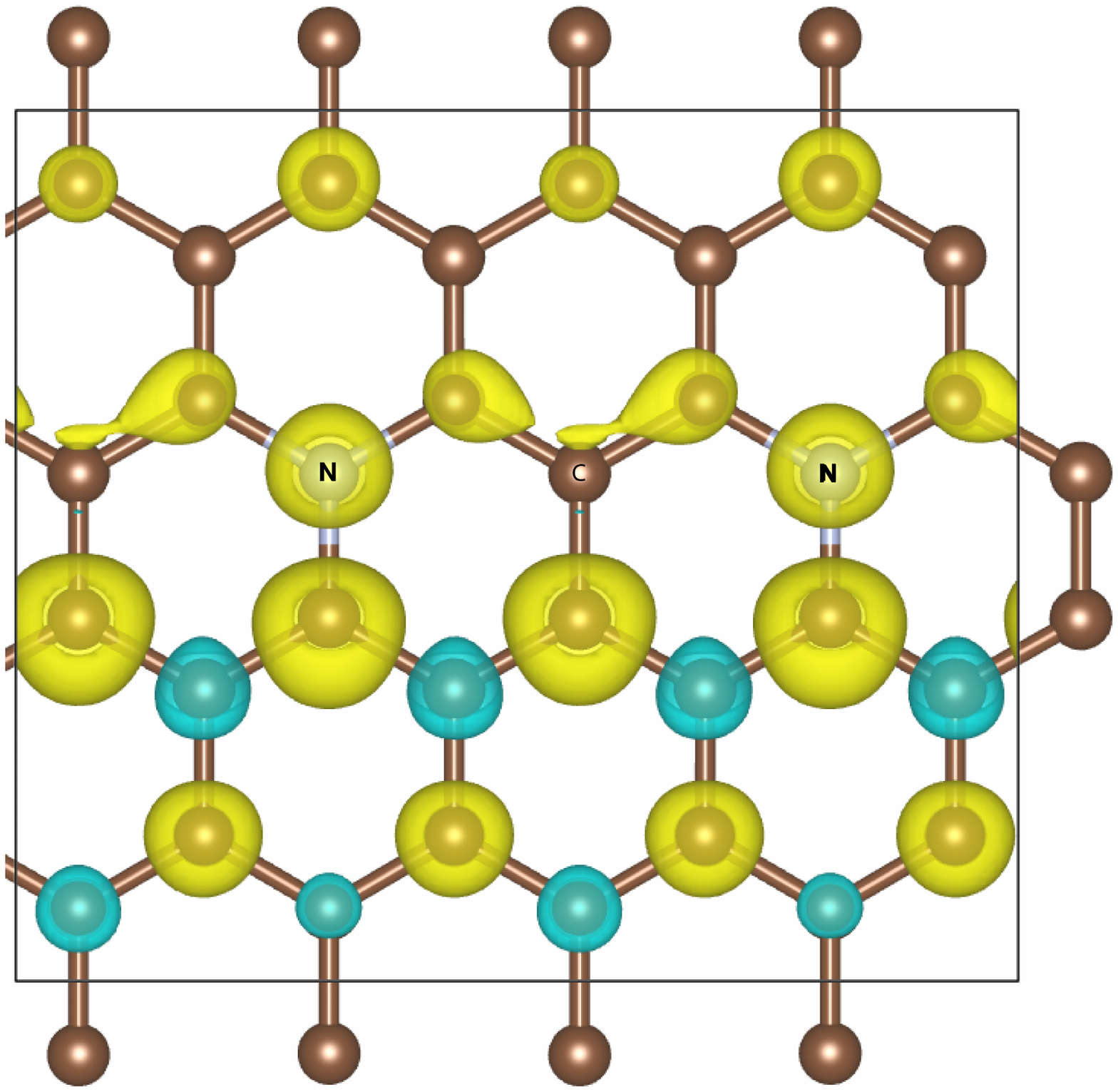}
	\caption{Simulation cell of the 6.2\% N-doped graphene calculated using the HSE06 functional. 
	The system has a small net magnetization moment at this level of theory, but not when a GGA or meta-GGA functional are used.
	The spin-up and spin-down densities are colored differently and the two N dopant atoms are marked. 
	Periodic images of some of the atoms are shown in addition to those within the simulation cell.
	 }
	\label{fgr:concentration}
\end{figure}

The free energy of reaction intermediates of ORR on NG is calculated using the 32-atom orthogonal simulation cell subject to periodic boundary conditions,
with either one  (gN1-G32) or two (gN2-G32)  graphitic N dopants included,
see figures \ref{fgr:dopedmodels} and \ref{fgr:concentration}. 
The resulting dopant concentration is 3.1\% and 6.2\%.
%
% -----------------------------------------  Figure 2 -----------------------------------------
% --------------------------------------------------------------------------------
%
% ------------------------------------------------------------------------ Figure 3 ------------------------------------------------------------------------
% ------------------------------------------------------------------------
%
This supercell is found to be large enough to obtain adsorption energy of ORR intermediates *O, *OH, and *OOH (see figure S1) 
unaffected by interaction with periodic images at the lower doping level. 
%An orthogonal simulation cell is used instead of the rhombic supercell used for the G32 benchmark. 
A $4 \times 4 \times 1$ Monkhorst-Pack\cite{monkhorst1976} $k$-point grid and 400 eV energy cutoff are found to give converged results (see table S2 and figure S2). SCAN requires a denser FFT integration grid than the other functionals as has been reported previously\cite{brandenburg2016}. To speed up the optimization of the atomic coordinates in the hybrid functional calculations, 
a $k$-point reduction scheme for the exact exchange step is used, reducing it to the $\Gamma$ point only.  
For a representative test system, the adsorption energy error introduced by this approximation is found to be $0.006-0.008$~eV (see Table S3). 
Final total energy results are obtained using the full $k$-point grid.

Zero point energy and vibrational entropy contributions are calculated from $\Gamma$ point phonon calculations within the harmonic approximation using four displacements of 0.015~\AA\ per degree of freedom. The graphene backbone is constrained in these calculations to reduce computational effort. For a representative test system and the PBE functional, the error introduced by this approximation is found to translate to an increase of \tcm\ by 0.04~V (see Tables S4 and S5). 
For phonon calculations using hybrid functionals, the same exact exchange $k$-point reduction scheme is used as for the energy
minimization.

Potential-dependent free energy diagrams at pH 0 are constructed using the TCM\cite{norskov2004,skulason2017}. 
From the free energy for each intermediate $x$, the reaction free energy is calculated as 
\begin{equation}
G^{\,x}_\text{ads}(U) = E^{\,x}_\text{el} + E_\text{ZPE}^{\,x} - T S^{\,x}_\text{vib} + n \text{e} U\, ,
\label{eq:gibbs}
\end{equation}
with $E^{x}_\text{el}$ being the total energy of intermediate $x$,   $E^{x}_\text{ZPE}$ the zero-point energy correction, 
%vibrational 
%  I hope you took the gas phase entropy into account for O2 !!! yes, the NIST-JANAF tabulated entropy - BK
$ T S^{x}_\text{vib}$ the entropy correction evaluated at $T = 298.15$~K, and $n \text{e} U$ describing the effect of the potential 
where $n$ is the number of electrons transferred during step $x$. Adsorption free energy values are calculated using gas-phase \ce{H2O} and \ce{H2} as reference, for which entropy contributions are obtained from the NIST-JANAF Thermochemical Tables\cite{chase1998}. 
By varying $U$ in \eqref{eq:gibbs}, an estimate of the onset potential is obtained as the point at which all reaction steps first become downhill. 
For an ideal ORR catalyst, this point would occur at 1.23~V. Non-ideal systems deviate from this and \tcm\ is obtained as the difference between 1.23~V and the calculated onset potential. See the Supporting Information for more details.

% -------------------

\subsection{Test of functionals: O-adatom on graphene}

\begin{figure}[h!]
	\centering
	\includegraphics[width=0.6\linewidth]{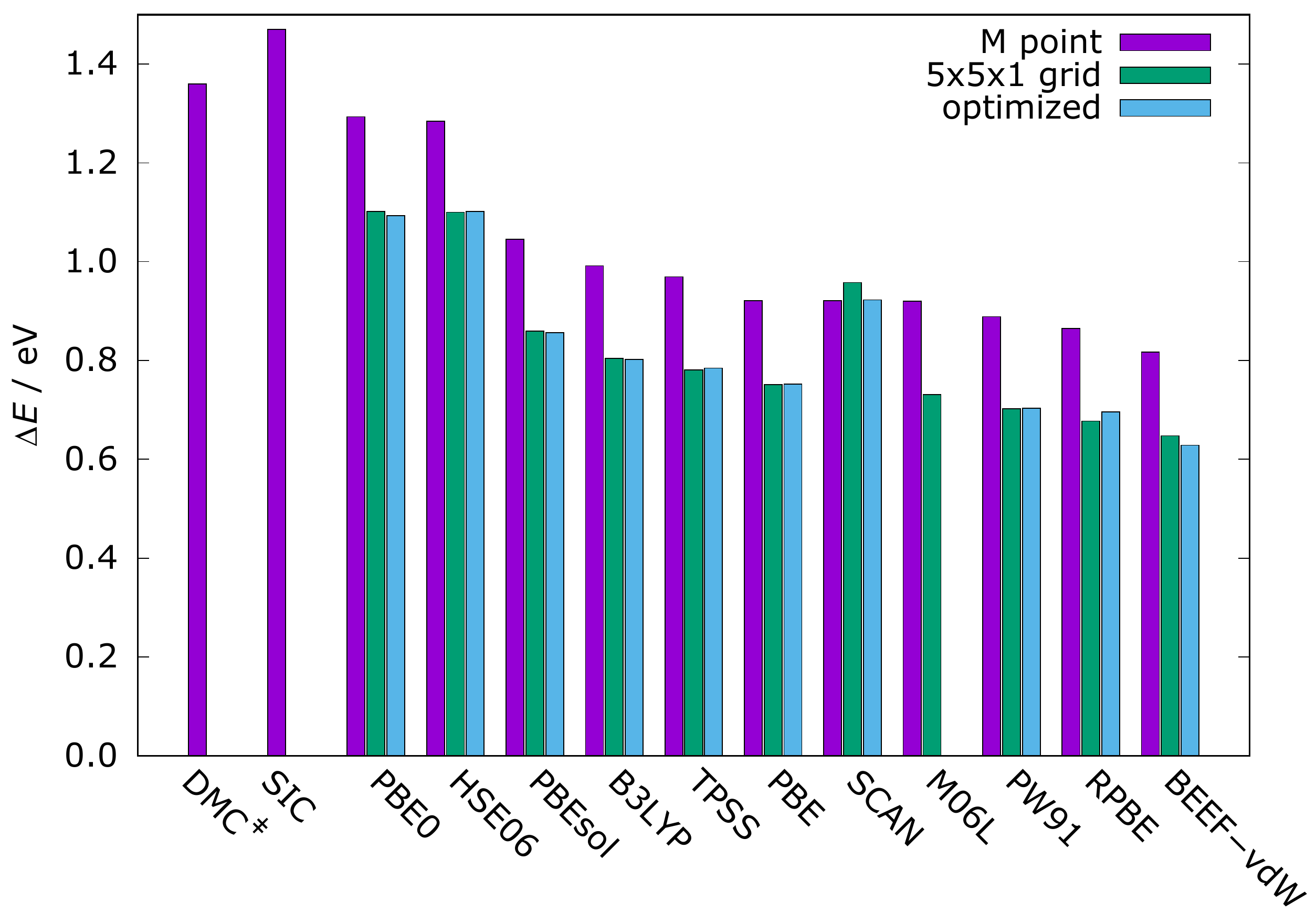}
% 	\caption{\added{Calculated energy difference $\Delta E$ = $E_{\text{tot}}^{\text{\,top}} - E_{\text{tot}}^{\text{\,bridge}}$ between
% 	         on-top and bridge sites of an O adatom on graphene using various DFT functionals, self-interaction corrected PBE and 
% 	         diffusion Monte Carlo. Purple bars show results obtained from \textit{M}(0.5, 0) point calculations for atom coordinates provided by Hsing \textit{et al.}\cite{hsing2012}
% 	         Green bars give $5 \times 5 \times 1$ \textit{\textbf{k}} grid calculations for the same atom coordinates.
% 	         Blue bars show results of $5 \times 5 \times 1$ \textit{\textbf{k}} grid calculations for atom coordinates optimized using the respective density functional.
% 	         $^\ddagger$Value obtained from \textit{M}(0.5, 0) point DMC calculation by Hsing \textit{et al}.\cite{hsing2012}}
% 	 }
	\caption{Calculated energy difference $\Delta E$ = $E_{\text{tot}}^{\text{\,top}} - E_{\text{tot}}^{\text{\,bridge}}$ between
	         on-top and bridge sites of an O adatom on graphene using various DFT functionals, self-interaction corrected PBE and 
	         diffusion Monte Carlo. Purple bars show results obtained from \textit{M}(0.5, 0) point calculations for atom coordinates provided by Hsing \textit{et al.}\cite{hsing2012}
	         Green bars give $5 \times 5 \times 1$ \textit{\textbf{k}} grid calculations for the same atom coordinates.
	         Blue bars show results of $5 \times 5 \times 1$ \textit{\textbf{k}} grid calculations for atom coordinates optimized using the respective density functional.
	         $^\ddagger$Value obtained from \textit{M}(0.5, 0) point DMC calculation by Hsing \textit{et al}.\cite{hsing2012}
	 }
	\label{fgr:hsing_mpoint}
\end{figure}

In order to test the accuracy of the various DFT functionals, a comparison is made with the published DMC results of Hsing {\it et al.} for the binding of an O adatom to graphene.\cite{hsing2012}
% A variation of the simulation strategy outlined there and by Janesko \textit{et al}.\cite{janesko2013} is adopted to produce the results 
The results are summarized in figure \ref{fgr:hsing_mpoint}. Here, the difference in total energy at the two sites for the O adatom, on-top and
bridge,  $\Delta E$ = $E_{\text{tot}}^{\text{\,top}} - E_{\text{tot}}^{\text{\,bridge}}$, is reported.  
 %while the absolute adsorption energy is not a focus since the O atom reference can introduce significant errors that are not so relevant for the present study.
%
% ------------------------------------------------------------------------ Figure 4 ------------------------------------------------------------------------

This measure ensures that the different density functionals are compared exclusively based on their description of the electronic structure of the graphene-oxygen system, eliminating contributions from the O atom reference that can be problematic in DFT calculations.\cite{klupfel2012,blochl2000}
The DMC results of Hsing \textit{et al.} were obtained for the special \textit{M}~=~(0.5, 0) $k$-point only. 
The dataset "\textit{M} point" in figure \ref{fgr:hsing_mpoint} is obtained using this $k$-point and the same atom coordinates as used by Hsing {\it et al.}

% \added{The hybrid functionals PBE0 and HSE06 give results that are closest to the DMC values, with deviations around 5\%.
% These functionals differ only by the inclusion of screening in HSE06. A 27\% lower value with respect to DMC is obtained with the B3LYP
% hybrid functional. These results are in agreement with the calculations of Janesko {\it et al.}\cite{janesko2013} 
% The GGA functionals PBE (32\% error), PW91 (35\%), RPBE (36\%) and BEEF-vdW (40\%) give results that differ significantly from the DMC results
% and have errors of similar magnitude the meta-GGA functionals, TPSS (29\%), SCAN (32\%) and M06L (32\%). The PBEsol functional produces a slightly smaller error of 23\% which is even lower than that of the B3LYP hybrid functional.}
The hybrid functionals PBE0 and HSE06 give results that are closest to the DMC values, with deviations around 5\%.
These functionals differ only by the inclusion of screening in HSE06. A 27\% lower value with respect to DMC is obtained with the B3LYP
hybrid functional. These results are in agreement with the calculations of Janesko {\it et al.}\cite{janesko2013} 
The GGA functionals PBE (32\% error), PW91 (35\%), RPBE (36\%) and BEEF-vdW (40\%) give results that differ significantly from the DMC results
and have errors of similar magnitude the meta-GGA functionals, TPSS (29\%), SCAN (32\%) and M06L (32\%). The PBEsol functional produces a slightly smaller error of 23\% which is even lower than that of the B3LYP hybrid functional.

The reason for the shortcoming of the GGA and meta-GGA functionals in this calculation can be traced to the self-interaction error that is
introduced in Kohn-Sham functionals that only depend on the total electron density.
The explicitly self-interaction corrected PBE functional gives an energy difference that is close to but even a bit 
higher than the DMC results, with 8\% deviation. 

The BEEF-vdW functional includes van der Waals interaction and several calculations were additionally carried out to assess the importance
of this contribution. 
Results obtained using the SCAN-rVV10\cite{peng2016} and DFT-D3BJ\cite{grimme2010,grimme2011} with PBE and HSE06\cite{moellmann2014} are given in table S6. 
The addition of van der Waals interaction does not change the calculated energy difference significantly.
%, which indicates that the contribution from dispersion in this benchmark system is negligible. 
The good performance of the self-interaction corrected PBE as well as the PBE0 and HSE06 hybrid functionals indicates that the dominant source of error in the GGA and meta-GGA calculations is due to the self-interaction. 

The DMC calculations are limited to just one $k$-point and do not include structure relaxation.
The effect of these two limitations is studied using the DFT functionals.
First, the influence of the $k$-point grid is investigated. To this end, $\Delta E$ is recalculated with the previously used density functionals on a 5$\times$5$\times$1 $k$-point grid on the same geometries used by Hsing and co-workers for DMC.  
While the $\Delta E$ values obtained this way cannot be directly compared to DMC, any changes in the relative sequence of functionals will reveal possible dependencies on the $k$-point sampling. Results of this test are shown in the dataset "5$\times$5$\times$1 grid" in figure \ref{fgr:hsing_mpoint}. 
The SIC method is omitted from hereon since the current implementation only supports single $k$-point calculations.
While the obtained $\Delta E$ values are overall lower by \textit{ca.} 0.2~eV compared to the \textit{M} point results, the relative ordering of 
results is consistent between \textit{M} point and full $k$-point grid calculations. The only outlier is SCAN, which, using the 5$\times$5$\times$1 $k$-point grid, produces a significantly higher $\Delta E$ value than all other meta-GGA and GGA functionals as well as B3LYP. Note however that SCAN requires a denser FFT grid than the other functionals for convergence of the \textit{M} point calculation. The numerical sensitivity of SCAN has been reported on in the past\cite{brandenburg2016}. The PBE result is identical to the value published by Hsing and \textit{et al.}\cite{hsing2012} and the overall trend is in agreement with the sequence published by Janesko and co-workers\cite{janesko2013}. Note that the values Janesko \textit{et al.} report are overall larger by 0.2--0.3~eV, which the group attributes to the use of an LCAO instead of a plane-wave basis set\cite{janesko2013}.

Secondly, the influence of the relaxation of the atomic configuration on the binding energy difference is investigated. So far, all $\Delta E$ calculations were performed using atomic configurations provided by Hsing and co-workers used for DMC calculations. Since different density functionals will produce different equilibrium lattice parameters, it is possible that this approach could introduce a form of lattice strain, leading to biased results. For this test, the atomic configurations and cell parameters of each system are therefore relaxed with each density functional using a $5\times5\times1$ $k$-point grid. Optimal graphite lattice parameters are obtained for each functional and are listed in table S1. The $\Delta E$ values obtained in this test are shown in the "optimized" dataset in figure \ref{fgr:hsing_mpoint}. 
%\added{The M06L functional is omitted from hereon due to convergence issues when relaxing the atomic configurations.}
The M06L functional is omitted from hereon due to convergence issues when relaxing the atomic configurations. No significant differences are found compared to column "$5\times5\times1$ grid", with only a small discrepancy observed for SCAN. These results suggest that using the same geometry throughout the \textit{M} point benchmark is inconsequential to the benchmark results. This conclusion also implies that for this material class, it is a suitable simulation approach to relax geometries at the GGA level and obtain accurate total energy values using a higher-level method. 

The BEEF-vdW functional  has an error estimate built in to indicate what range of values can
be expected from any reasonable parametrization of the GGA functional form. 
The one standard deviation error bar obtained with an ensemble size of 2000 for the binding energy difference is quite large, 0.2 eV,
corresponding to $\pm$~30~\%. 

% ----------------------------------------------------------------------------------------------------------------------------------------------------------------

\section{Results}

Results obtained for periodic models of doped graphene are presented first and then results obtained using finite, flake-like models.
%\subsection{Thermochemical Overpotential \tcm\ on N-doped Graphene}

\subsection{A. Periodic with 3.1\% and 6.2\% doping}

The ORR overpotential for the periodic models of NG are calculated with the same set of density functionals used in the test against the DMC 
results for the O adatom on graphene. Given the good performance there, HSE06 is herein used as the best estimate for the overpotential 
and as the baseline for comparison.
The periodic model systems of NG contain 32 atoms in the simulation cell and a dopant concentration of 3.1\% (gN1-G32p) as illustrated in figure \ref{fgr:dopedmodels} 
and 6.2\% (gN2-G32p) as illustrated in figure \ref{fgr:concentration}.
These concentration values are within the 1 to 10\% range
%usual range 
reported for experimentally studied materials\cite{wang2012}.
The calculated values of the adsorption energy for the various ORR intermediates are converged with respect to system size in the low concentration model (see figure S1). 

\begin{figure}[h!]
	\centering
	\includegraphics[width=0.9\linewidth]{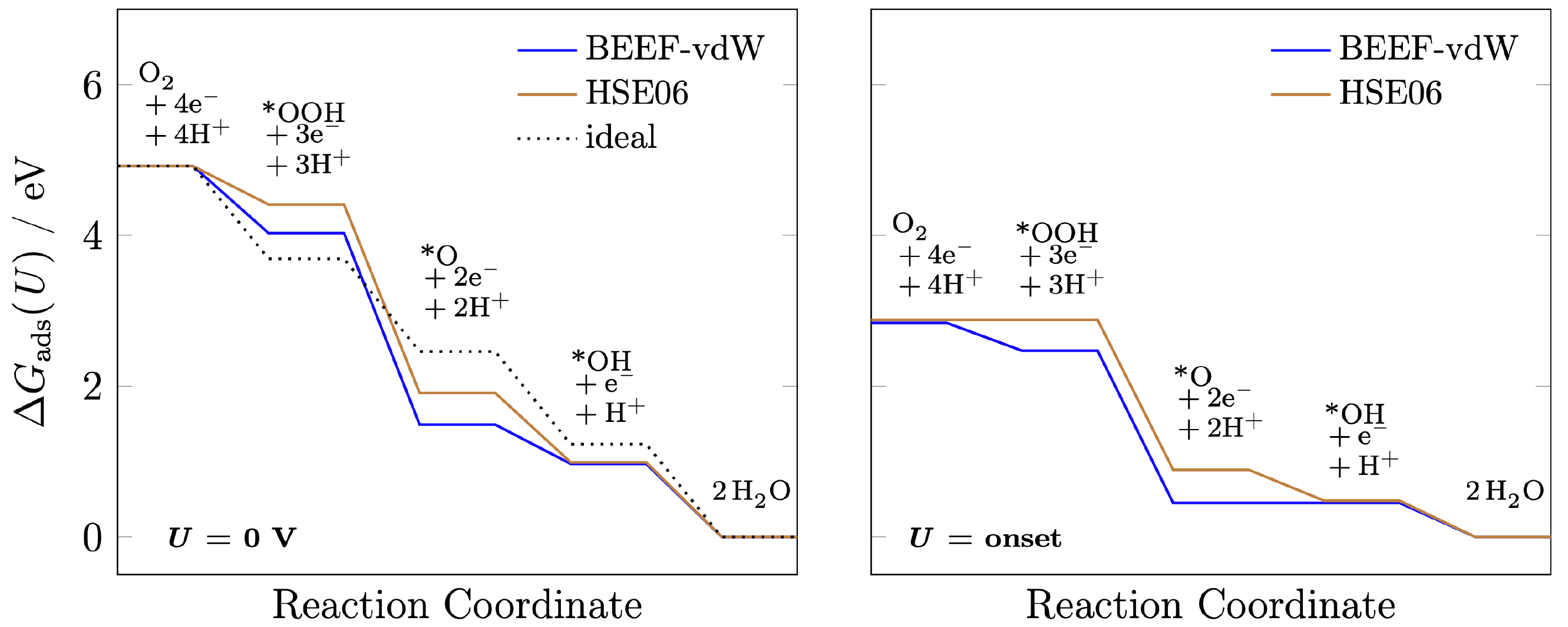}
	\caption{
	Calculated free energy for the oxygen reduction reaction on graphene with 6.2\% N-dopant concentration.
	Left:  at zero voltage.  Right: at voltage giving as flat free energy profile as possible without any step
	being uphill.  For the BEEF-vdW this voltage is 0.52 V corresponding to \tcm\ = 0.71 V 
	and the reduction of *O is the potential determining step
	(as for the other GGA functionals),
	while for the HSE06 this voltage is 0.51 V  corresponding to \tcm\ = 0.72 V 
	and the reduction of O$_2$ to form *OOH is the potential determining step.
	}
	\label{fgr:freeenergy}
\end{figure}

\begin{figure}[h!]
	\centering
	\includegraphics[width=0.5\linewidth]{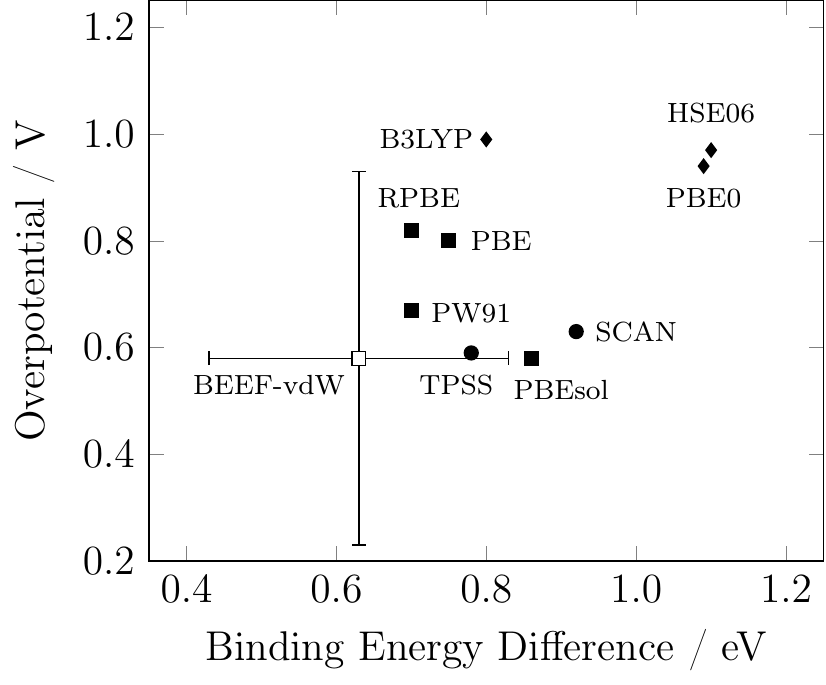}
% 	\caption{\added{Thermochemical overpotentials \tcm\ for 3.1\% N-doped graphene calculated using various DFT functionals 
% 	vs. the binding energy difference for the O adatom, $\Delta E$, from the "optimized" dataset shown in figure \ref{fgr:hsing_mpoint}.
%         The general trend is that functionals giving a more accurate estimate of $\Delta E$ give higher values of the onset potential \tcm\ .
% 	GGA: squares;  meta-GGA: circles; hybrid functionals: diamonds.}}
	\caption{Thermochemical overpotentials \tcm\ for 3.1\% N-doped graphene calculated using various DFT functionals 
	vs. the binding energy difference for the O adatom, $\Delta E$, from the "optimized" dataset shown in figure \ref{fgr:hsing_mpoint}.
        The general trend is that functionals giving a more accurate estimate of $\Delta E$ give higher values of the onset potential \tcm\ .
	GGA: squares;  meta-GGA: circles; hybrid functionals: diamonds.}
	\label{fgr:opvsmpoint}
\end{figure}

The calculated free energy of the intermediates in the \ce{4 e-} ORR reaction path is shown in figure S3 for the lower dopant concentration and
in figure \ref{fgr:freeenergy} for the higher dopant concentration. 
The onset potential, U$_o$, that makes the potential determining step flat in free energy is determined
for each case, and from that the overpotential is estimated as  \tcm\ = 1.23 V - U$_o$. 
For the lower dopant concentration, a large range of values is obtained and a correlation between the value of the estimated
overpotential and the binding energy difference of the O adatom is noted.  This is illustrated in 
figure \ref{fgr:opvsmpoint} which shows the calculated \tcm\ values for each functional as a function of the $\Delta E$ values from the 'optimized' dataset shown in figure \ref{fgr:hsing_mpoint}.
%
% ----------------------------------------------  Figure 5 ---------------------------------------------------------
% ----------------------------------------------------------------------------------------

% ----------------------------------------------  Figure 6 ---------------------------------------------------------
% ----------------------------------------------------------------------------------------
%
The onset potential is limited by the formation of *OOH in all cases for the lower dopant concentration.
% (see Figure S3 for free energy diagrams). 
The hybrid functionals, PBE0 and HSE06, produce the highest \tcm\ values of \textit{ca.}~1.0~V. B3LYP produces a \tcm\ similar to the HSE06 reference value despite showing larger deviations from the DMC results on the calculations of the *O binding energy difference. This is a result of B3LYP producing *OH and *OOH free energy values similar to PBE0 and HSE06 while a larger deviation is observed  for the free energy of the *O intermediate. Since ORR onset potential on this particular model system is limited by the *OOH formation step, the \tcm\ obtained with B3LYP is in good agreement with PBE0 and HSE06.
%; however, differences will be more pronounced for systems that are more sensitive to changes of the free energy of the *O intermediate. Examples of such systems will be discussed below. 

The results obtained with PBE, RPBE, PW91, TPSS, and SCAN functionals are found to be similar, analogous to the calculations of the *O binding energy difference. Compared to PBE0/HSE06, they underestimate \tcm\ by 0.2--0.4~V. %\added{The PBEsol functional underestimates \tcm\ in a similar way despite better performance in the benchmark compared to the other (meta-)GGA functional. PBEsol is not optimized for surface adsorption calculations and its poor performance in this regard is well documented.\cite{wellendorff_benchmark_2015}}
The PBEsol functional underestimates \tcm\ in a similar way despite better performance in the benchmark compared to the other (meta-)GGA functional. PBEsol is not optimized for surface adsorption calculations and its poor performance in this regard is well documented.\cite{wellendorff_benchmark_2015}
The BEEF-vdW functional underestimates \tcm\ by \textit{ca.}~0.4~V. The vertical error bars for BEEF-vdW in figure \ref{fgr:opvsmpoint} indicate the uncertainty of \tcm\ based on the standard deviation of the free energy values used to calculate \tcm, obtained from an ensemble size of 2000 functionals. 
The error bars span a wide range of values, between 0.2--1.0~V. Note that PBE-D3 gives an almost identical \tcm\ value as BEEF-vdW, indicating that the observed difference between BEEF-vdW and PBE is likely due to  the added dispersion energy. Based on these results and the earlier benchmark, inclusion of exact exchange in hybrid functionals appears to be necessary to obtain accurate TCM results for graphene-based materials. Care should be taken in the choice of functional if correct description of the *O intermediate is crucial; there, the B3LYP functional is less accurate. 
The \tcm\ value of \textit{ca.}~1.0~V obtained with the HSE06 functional is significantly higher than values published previously, as discussed
above in the introduction. 
%Studt calculated a \tcm\ of 0.72~V for a model system with a graphitic dopant concentration of 6.3~\% using the BEEF-vdW functional\cite{studt2013}. Sinthika an co-workers extrapolated an ideal NG catalyst with a graphitic dopant concentration of 6.0~\% using a descriptor approach and the PBE functional, for which they calculated a \tcm\ of 0.48~V\cite{sinthika2018}. This result would indicate that the associative 4 \ce{e^-} mechanism is not favorable for NG with N dopants in the basal plane.

%However, further comparison of the obtained results with literature studies is limited because the model chosen for this benchmark is less complex than some of those used in application-focused studies. To address this, two important factors deemed crucial in past studies are explored in more detail, namely inclusion of water\cite{okamoto2009,reda2018} and dopant concentration\cite{sinthika2018,reda2018}.

% ------------------------------------------------------------------------------

%\subsection{Influence of the Dopant Concentration}

%Dependence of \tcm\ on the dopant concentration is investigated for NG with and without solvation contributions. To this end, the dopant concentration is doubled from 3.1~\% to 6.2~\%. Free energy diagrams for this model are presented in the Supporting Figure 5. 

When the dopant concentration is doubled to 6.2\% the electronic structure of NG calculated with HSE06 shows interesting features.
%Firstly, it is of interest to analyze the electronic structure of these NG models in more detail since it has recently been shown that c
The calculated spin density is illustrated in figure \ref{fgr:concentration} 
and shows non-zero net magnetization as well. This is in agreement with experiments which show that certain dopant cluster arrangements lead to a ferromagnetic ground state 
%In fact, unlike for the single-doped system, the model with higher dopant concentration shows 
with non-zero magnetization.\cite{blonski2017}
However, in the GGA calculations, no net magnetic moment is obtained. This result indicates that GGA functionals fail to describe some fundamental aspects of the electronic structure of this material.

Using the HSE06 functional, the binding strength of all ORR intermediates increases when the dopant concentration is doubled, with the largest increase obtained for *O. This leads to a decrease of \tcm\ to 0.72~V, with *OOH reduction still remaining the potential determining step. 
%Studt found a similar value for an identical model system using BEEF-vdW\cite{studt2013}. Note however that Studt found *O reduction to be the PDS, likely as a direct result of the *O overbinding issue highlighted for BEEF-vdW in the benchmark. This agreement is therefore coincidental.
In the calculations using PBE and BEEF-vdW, the potential determining step is the reduction of *O.
The hybrid functionals and GGA functionals, therefore, give quite different trends for ORR when the dopant concentration is 6.2\%
even though the estimated overpotential is quite similar.

% ------------------------------------------------------------------------------------------------------------------------------

\subsection{B. Finite flake models}
 
Calculated  \tcm\ for finite model systems similar to those used in several previous studies\cite{sidik2006,zhang2011,zhang2012,jiao2014}
are shown in figure 6.
In particular, the calculations are performed for a 54-atomic NG flake akin to the model used by Jiao and co-workers\cite{jiao2014} (see figure S6)
where the dopant concentration is 2.4~\%. Using the PBE0 functional, a \tcm\ value of 0.62~V is obtained. 
%which is further reduced to 0.46~V once solvation contributions are considered in the same fashion as before. 
These results are noteworthy because the \tcm\ values obtained this way are significantly lower than results for the periodic model systems at 3.1~\% dopant concentration (0.62~V \textit{vs.} 0.94~V with PBE0). Results from the periodic and molecular flake models are, therefore, not consistent since in the periodic case, a trend was established relating increasing dopant concentration to lower \tcm\ values. Furthermore, the size of the periodic model is shown to be converged with regards to the adsorption energy values of ORR intermediates as shown in the figure S1. Reducing the dopant concentration of the periodic model system does not change the obtained \tcm\ value. In the following, two potential causes of this deviation between molecular and periodic model systems are investigated in more detail, finite size effects and geometric distortion of the flake models.

\begin{figure}[h!]
	\centering
	\includegraphics[width=0.9\linewidth]{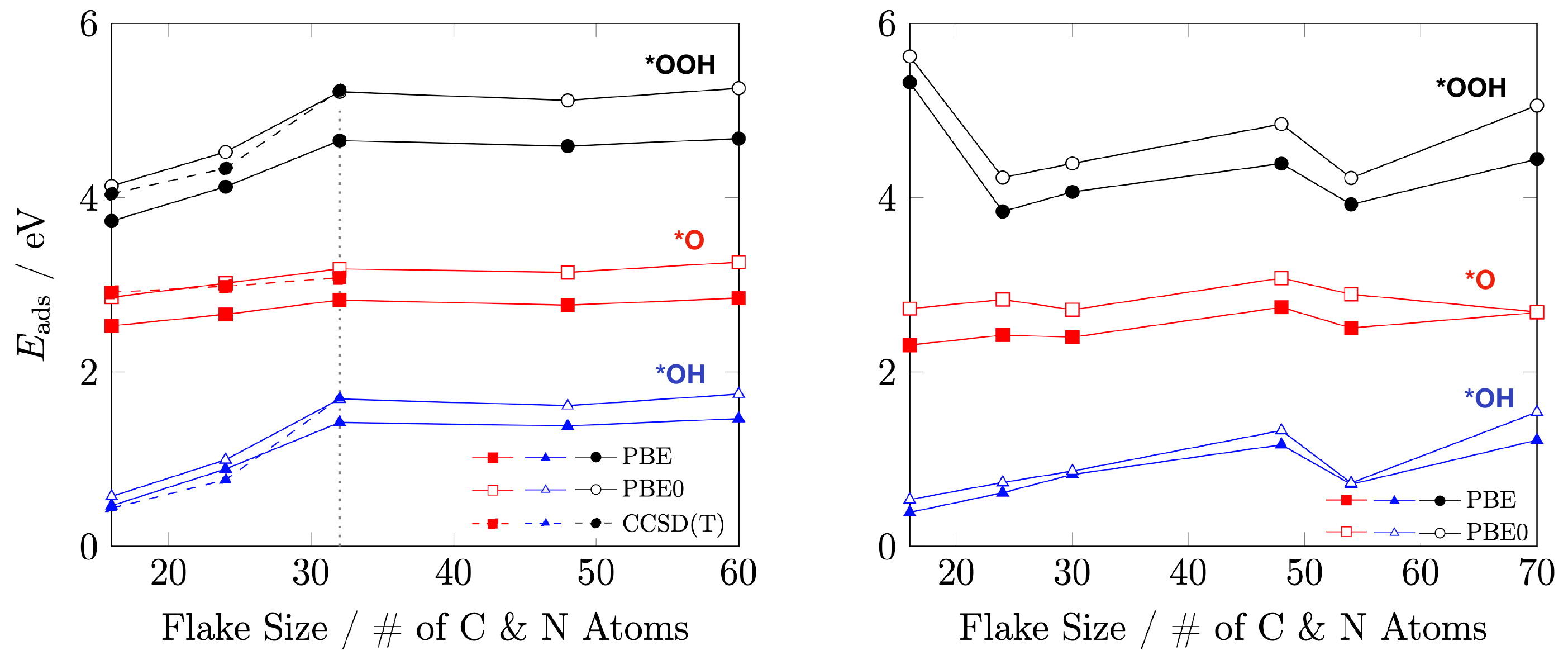}
	\caption{Adsorption energy of the *O, *OH and *OOH for flakes containing one N atom as a function of the total number of atoms in the flake.
%	 using the PBE and PBE0 functionals. 
	Left: Flakes shaped as the simulation cell used in the periodic calculations but capped with hydrogen atoms.
	    The atomic coordinates are fixed as in the relaxed periodic configurations except that the capping hydrogen atoms are allowed to relax.  
	    High-level, wave function based DLPNO-CCSD(T) calculations show that the PBE0 functional gives more accurate results than PBE.
	 Right: Round or diamond-shaped flakes where all atom coordinates are allowed to change as the energy is minimized.
%	 $^{\dagger}$: DLPNO-CCSD(T) shortened to CCSD(T) for brevity.
          }
	\label{fgr:flakes}
\end{figure}

To study the influence of finite size effects in more detail, adsorption energy calculations of ORR intermediates are performed using the PBE and PBE0 functionals on a set of hydrogen-terminated models generated from the periodic size convergence study shown in figure 6. These models therefore retain the atomic configurations of the periodic structures. Only the hydrogen atoms are allowed to relax during the calculations. Figure \ref{fgr:flakes} \textbf{a} shows the obtained adsorption energy trends as a function of the flake size.
%
%   ---------------------------------------------  Figure 7 ------------------------------------------------------
%   ----------------------------------------------------------------------------------------------------------------
%
Since all the flake models contain one N atom, the dopant concentration decreases as a function of the flake size. 
In order to test the accuracy of the DFT functionals, high level wave function based calculations are performed and treated as benchmark results.
DLPNO-CCSD(T) calculations are performed for flakes of up to 32 atoms. The results confirm the PBE0 functional to be most accurate for this application. The flakes in this set show similar convergence behavior to the periodic models but admolecules are overall less strongly bound compared to the periodic calculations with HSE06. This decreased binding strength results in a very high \tcm\ value of \textit{ca.}~1.7~V in case of the converged 32-atomic model. It is suggested that this deviation is a result of finite size effects modifying the adsorption energetics. Note that the different choice of basis set can influence this comparison as well but the difference between LCAO and PW showcased by Janesko and co-workers\cite{janesko2013} is not large enough to fully account for the \textit{ca.}~1~eV increase in adsorption energy of the *OOH intermediate on the flake model compared to the periodic calculation.

The second set of calculations is performed on round and diamond-shaped molecular flakes of increasing size. They are not generated from periodic models and the atomic coordinates are not constrained in any way. These models also contain only one dopant atom. Strong deformation of the flakes is observed during energy minimization, see figure S. Adsorption energy values as a function of the flake size are shown in figure \ref{fgr:flakes} \textbf{b}. Unlike for the flake models derived from periodic configurations of the atoms, no obvious convergence trend is observed with regards to the flake size. Adsorption energy values fluctuate by as much as 0.8~eV up to a flake size of 70 C \& N atoms, at which point hybrid calculations start to become too computationally demanding. Notably, the 54 atom flake, which was used to calculate the \tcm\ values at the start of this section, shows strong binding of ORR intermediates compared to the flakes derived from periodic models and therefore produces a small \tcm. Overall, the results show that geometric distortions of the flakes affect the calculated adsorption energy in unpredictable ways. 
%A recent study by Xie and co-workers suggests that geometric distortions in N-doped graphene can break the scaling relations, which could partly explain this erratic behavior\cite{xie2018}.
%
%  -------------------------------------------   Figure 8 -------------------------------------------------
\begin{figure}[h!]
	\centering
	\includegraphics[width=\linewidth]{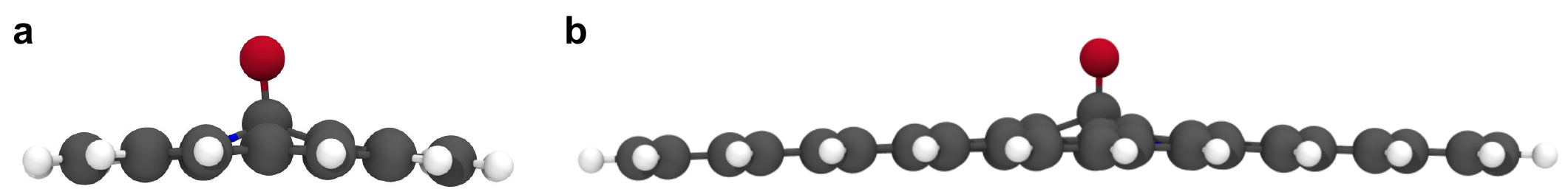}
	%\caption{\added{Side view of an oxygen atom adsorbed on molecular flake model systems with \textbf{a} 24 and \textbf{b} 70 C and N atoms. Convex distortion of the flake is most obvious for the small model systems but can still be observed at the large model size (\textbf{b)}.}}
	\caption{Side view of an oxygen atom adsorbed on molecular flake model systems with \textbf{a} 24 and \textbf{b} 70 C and N atoms. Convex distortion of the flake is most obvious for the small model systems but can still be observed at the large model size (\textbf{b)}.}
	\label{fgr:distortion}
\end{figure}
%  ---------------------------------------------------------------------------------
%
Free energy diagrams are shown in figures S7 and S8. For this comparison, only the electronic ground state energy values are considered since ZPE and entropic contributions do not change the trends. The \tcm\ trends in figure \ref{fgr:flakes_OP} directly correspond to changes of the *OOH adsorption energy values in figure \ref{fgr:flakes}. This relationship is a result of *OOH formation being the potential determining step in all the models. It is noteworthy that for the round and diamond-shaped systems, which are not generated from periodic systems, the flakes with 24 and 54 C and N atoms give rise to the lowest \tcm. These two models are round whereas the others are diamond shaped, see figure S6. The shape of the model therefore seems to influence its thermochemistry significantly. This does not appear to be the case for the models generated from periodic models, even though the periodic models presented in the size convergence study (see figure S1) also vary in their aspect ratio.

%  ---------------------------------------------------------------------------------------------------------------------------------------------------------------------

Lastly, \tcm\ values are calculated for these two sets of finite NG models and the results are shown in Figure \ref{fgr:flakes_OP}.
%
%  -------------------------------------------   Figure 9 -------------------------------------------------
\begin{figure}[h!]
	\centering
	\includegraphics[width=0.5\linewidth]{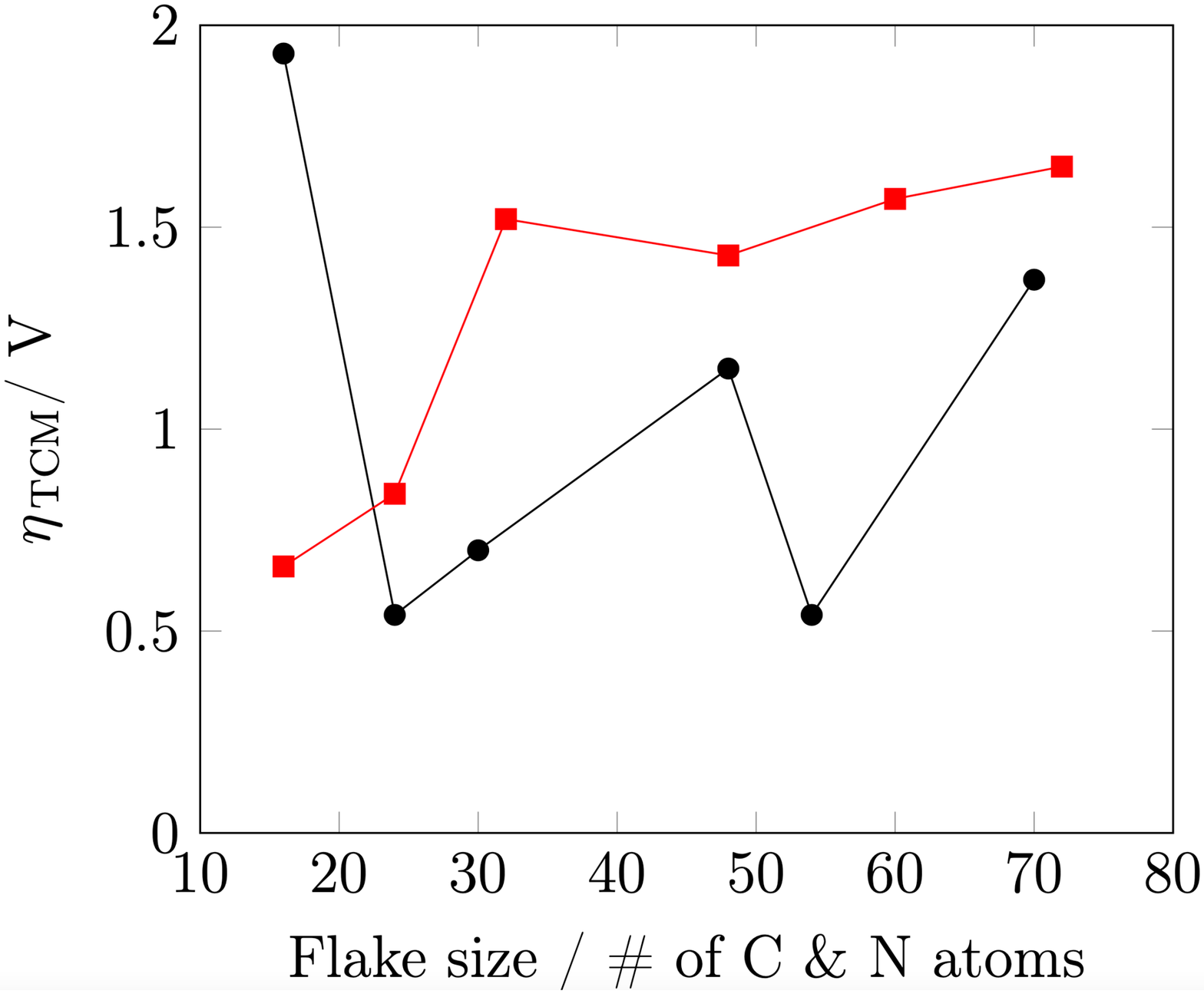}
	\caption{Estimated \tcm\ for round and diamond-shaped flakes allowed to fully relax during calculations and flakes with fixed geometry generated from (relaxed) periodic models that are capped with hydrogen atoms.}
	\label{fgr:flakes_OP}
\end{figure}
%  ---------------------------------------------------------------------------------
%
Free energy diagrams are shown in figures S7 and S8. For this comparison, only the electronic ground state energy values are considered since ZPE and entropic contributions do not change the trends. The \tcm\ trends in figure \ref{fgr:flakes_OP} directly correspond to changes of the *OOH adsorption energy values in figure \ref{fgr:flakes}. This relationship is a result of *OOH formation being the potential determining step in all the models. It is noteworthy that for the round and diamond-shaped systems, which are not generated from periodic systems, the flakes with 24 and 54 C and N atoms give rise to the lowest \tcm. These two models are round whereas the others are diamond shaped, see figure S6. The shape of the model therefore seems to influence its thermochemistry significantly. This does not appear to be the case for the models generated from periodic models, even though the periodic models presented in the size convergence study (see figure S1) also vary in their aspect ratio.

%  ---------------------------------------------------------------------------------------------------------------------------------------------------------------------

\section{Discussion}

In this study, the accuracy of various density functionals is assessed for adsorption of oxygen-containing adspecies on graphene-based materials. 
The thermochemical overpotentials \tcm\ for the associative \ce{4 e-} ORR mechanism on a periodic NG model system are calculated with  GGA, meta-GGA and hybrid density functionals. Results indicate that GGA functionals underestimate \tcm\ by up to 0.4~V when the dopant concentration is low, 3.1\%. The hybrid PBE0 and HSE06 functionals
which have been shown to give more accurate estimates of the binding of an O adatom on pure graphene as compared to DMC results,
give a larger estimate of the overpotential, of 1.0 V.
The meta-GGA functionals, TPSS and SCAN, give values of the overpotential that are similar to the ones obtained from GGA functionals.
Bayesian error estimation with the BEEF-vdW functional produces a large uncertainty of $\Delta$\tcm~$\approx$~0.8~V indicating that
the GGA functional form is unreliable. It should be noted, however, that this error estimate does not take into account systematic errors
in the GGA functional form, such as the self-interaction error, which is shown here to be responsible for the shortcomings of GGA in
calculations of the O adatom binding on graphene. 
The *OOH intermediate can be further stabilized by increasing the dopant concentration, which decreases \tcm\ obtained by HSE06 to 0.72~V. This trend is opposite to the GGA results where the reduction of *O becomes the potential determining step. 
The trends obtained with GGA functionals can therefore be quite different from the trends obtained with hybrid functionals for this reaction.

% -----   Solvation
It is clear from these studies that hybrid functionals give more accurate results than GGA or meta-GGA functionals. 
The main uncertainty that remains is the effect of solvation, \textit{i.e.} the presence of the aqueous electrolyte and in particular hydrogen
bonding of the admolecules with water molecules.
Various different estimates of the shift in free energy of ORR intermediates due to the presence of water have been presented in the
literature.
The classical dynamics simulations of Yu {\it et al.}\cite{yu2011} give a lowering of the binding free energy of all the intermediates by about 0.5 eV,
but the calculations of Reda {\it et al.}\cite{reda2018} using ice layers indicate a lowering of 0.18 eV  for *OOH and *OH and about twice as much for *O.
Such contributions from solvation can change the estimated  \tcm\ significantly.
If the free energy of *OOH is lowered by 0.2 eV, the \tcm\ obtained from calculations where the formation of *OOH is the potential determining
step is reduced by 0.2 V. For example, the  \tcm\ obtained from the HSE06 functional becomes \textit{ca.}~0.8~V for the 3.1\% doping concentration. 

%Using the HSE06 function, the stabilization of the ORR intermediates resulting from explicit water is found to be consistent with published results\cite{reda2018} at the GGA level, see Supporting Table 7. This quantity therefore appears robust against the choice of functional, thus allowing to be used as a post-correction to the \tcm\ values shown in Figure \ref{fgr:opvsmpoint}. Table \ref{tbl:solvation} contrasts the \tcm\ values with and without solvation contribution. 
Similarly, for the higher dopant concentration of 6.2~\%, 
%If the influence of water is taken into account in the same fashion as in section \ref{s:water}, 
the \tcm\ value obtained with HSE06 is reduced from 0.72~V to 0.54~V, which is in the same range as commercial Pt catalysts. 
%This is much lower than the values found by Reda and co-workers\cite{reda2018} who, similar to Studt, found *O reduction to be the PDS using BEEF-vdW. Similarly, this result is also in disagreement with a PBE-based AIMD study by Okamoto which includes water explicitly\cite{okamoto2009}. Like Studt and Reda \textit{et al.}, Okamoto found the *O reduction step to become limiting with increasing dopant concentration. The present results, therefore, show that GGA functionals can produce different overall trends compared to hybrid functionals for this class of materials.
%
%Since for all functionals, the limiting process is *OOH reduction, the solvation stabilization for the *OOH admolecule, $\Delta E_{\text{solv}}^\text{OOH} = -0.18$~eV\cite{reda2018}, directly translates into a decrease of \tcm\ by 0.18~V. 
Figure S4 shows free energy diagrams with and without solvation contributions for the lower dopant concentration,
where the solvation effect estimated by Reda {\it et al.} is applied.
%This prediction is lower than the BEEF-vdW-based estimates of 0.78, 0.94, and 1.14~V published by Reda \textit{et al.} for NG with a higher dopant concentration of 6.3~\% at adsorbate coverages of $\theta$ = 1.0, 0.5, and 0.33, respectively, in the presence of explicit solvation\cite{reda2018}.
%
%The results presented here can serve as a first estimate regarding influence of solvation on the adsorption energy. Entropic contributions are not considered. 

Ultimately, extensive sampling of solvent configurations around the adsorbate taking into account the electrochemical environment
is required to refine the calculated estimates of the overpotential in ORR. Furthermore, the effect of free energy barriers in the elementary
steps may need to be taken into account instead of just the thermodynamics of the intermediates.
The simulations of electrochemical systems is a challenging but rapidly advancing field and further development of the simulation methodology
will in the future make more accurate estimates of the electrochemical overpotential possible. 
%This can be achieved for example molecular dynamics simulations combined with DFT or a QM/MM scheme where DFT calculations for part of the system are coupled to a description of the water phase using a potential energy function.

% ----  flakes

Lastly, doped molecular flakes are investigated. The \tcm\ from flake models which retain the atomic configuration of a periodic model system is larger than 1.0~V, indicating that quantum size effects strongly decrease adsorption strength for all ORR intermediates. However, for a set of typical round and diamond-shaped flakes, strong geometric distortion during relaxation is observed. This leads to erratic jumps in the adsorption energetics of ORR intermediates for different flake sizes and shapes.

Further research using hybrid DFT functionals is required to explore the debated adsorption mode of \ce{O2} and kinetic aspects of ORR on NG. Edge-effects, the influence of geometric distortions like the Stone-Wales defect, and the activity of B-doped graphene models should be re-investigated using periodic model systems and the HSE06 functional to confirm or correct previously published trends. Finally, solvation contributions, which are found to affect adsorption free energy values significantly, need to be obtained using rigorous sampling of solvent configurations and the explicit electrode potential needs to be taken into account.

%%%%%%%%%%%%%%%%%%%%%%%%%%%%%%%%%%%%%%%%%%%%%%%%%%%%%%%%%%%%%%%%%%%%%
%\clearpage

\begin{acknowledgement}

This work was supported by the Icelandic Research Fund.
BK thanks the University of Iceland Research Fund for a doctoral fellowship.
Dr. Cheng-Rong Hsing and Prof. Mei-Yin Chou are acknowledged for providing graphene geometries
and Dr. Yan Jiao is acknowledged for providing graphene flake geometries used in their respective studies.
Dr. Maxime Van den Bossche is thanked for fruitful discussions. 
%The Icelandic Research Fund (grant no. 174582-052) is acknowledged. 
The calculations were performed at the Icelandic Research High Performance Computing center at the University of Iceland.

\end{acknowledgement}

%%%%%%%%%%%%%%%%%%%%%%%%%%%%%%%%%%%%%%%%%%%%%%%%%%%%%%%%%%%%%%%%%%%%%

\begin{suppinfo}

Data on adsorption energy calculations, vibrational frequencies and resulting free energy corrections, as well as a list of all adsorption configurations for ORR intermediates can be found in the Supporting Information.

\end{suppinfo}

%%%%%%%%%%%%%%%%%%%%%%%%%%%%%%%%%%%%%%%%%%%%%%%%%%%%%%%%%%%%%%%%%%%%%

\bibliography{benchmark}

\cleardoublepage



\section*{Supplementary material}

# Supporting Information for:

# Accuracy Assessment of Density Functionals for the Oxygen Reduction Reaction on Doped Graphenes

Björn Kirchhoff,[†,‡] Aleksei Ivanov,[†] Egill Skúlason,[¶] Timo Jacob,[‡,§,∥] Donato Fantauzzi,[†] and Hannes Jónsson[*,†]

†Science Institute and Faculty of Physical Sciences, University of Iceland, VR-III, Hjarðarhagi 2, 107 Reykjavík, Iceland

‡Institute of Electrochemistry, Ulm University, Albert-Einstein-Allee 47, 89081 Ulm, Germany

¶Science Institute and Faculty of Industrial Engineering, Mechanical Engineering and Computer Science, University of Iceland, VR-III, Hjarðarhagi 2, 107 Reykjavík, Iceland

§Helmholtz-Institute Ulm (HIU) Electrochemical Energy Storage, Helmholtz-Straße 16, 89081 Ulm, Germany

∥Karlsruhe Institute of Technology (KIT), P.O. Box 3640, 76021 Karlsruhe, Germany

E-mail: hj@hi.is

Table 1: C–C distances ($d_{\text{C–C}}$) and interlayer distances $d_{\text{layers}}$ of P6$_3$/mmc graphite bulk cells optimized at the respective level of theory compared to experimental values.

| Position | $d_{\text{C–C}}$ / Å | $d_{\text{layers}}$ / Å |
|---:|:---:|:---:|
| Experimental[1,2] | 1.420 | 3.335 |
| PBE | 1.424 | 3.929 |
| PBE-D3 | 1.423 | 3.343 |
| BEEF-vdW | 1.423 | 3.630 |
| TPSS | 1.423 | 4.380 |
| SCAN | 1.416 | 3.441 |
| SCAN-rVV10 | 1.416 | 3.343 |
| PBE0 | 1.414 | 4.520 |
| HSE06 | 1.415 | 3.766 |
| HSE06-D3 | 1.414 | 3.351 |
| B3LYP | 1.419 | 3.922 |

# 1 Optimized cell parameters

# 2 Supercell size convergence study

Supercell size convergence for Nitrogen-doped graphene (NG, Figure 1). Supercells always included one graphitic dopant; the dopant concentration therefore is not static. All supercells were converged with respect to the $\boldsymbol{k}$ grid before evaluation. Note that HSE06 values are obtained using the $\boldsymbol{k}$ grid reduction scheme for the exact exchange (EE) step outlined in the main manuscript. Also, adsorption energy values given here may vary from the final reported values since they were obtained with less strict precision parameters. The NG model seems to require a 32 atomic cell in particular for converged *OH results.

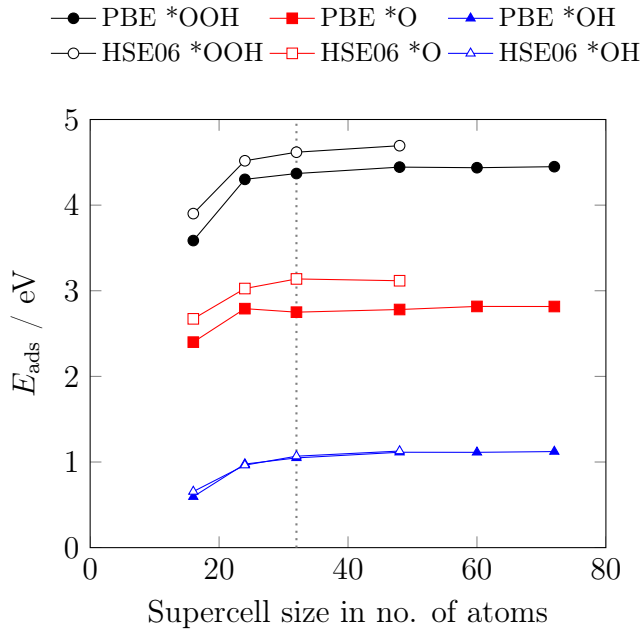

Figure 1: Supercell size convergence study for cells containing 1 graphitic Nitrogen atom at the PBE and HSE06 level, with the latter using the $k$ point reduction scheme outlined in main MS.

## 3 $k$ Grid convergence of $\eta_{\text{TCM}}$

We investigate the dependence of the thermodynamic reaction overpotentials $\eta_{\text{TCM}}$ with respect to the $k$ point grid for the PBE, SCAN, and HSE06 density functional approximations. The calculation parameters are the same as for the final reported values in the main manuscript. HSE06 values reported here used the full $k$ grid for the HF step. Table 2 reports values for NG. Note that $k$ grid convergence with SCAN difficult, showing decreasing $\eta_{\text{TCM}}$

Table 2: $k$ grid convergence study for thermodynamic reaction overpotentials $\eta_{\text{TCM}}$ on gN1-G32 in V.

| | $\eta_{\text{TCM}}$ | | | |
| --- | --- | --- | --- | --- |
| Functional | $3 \times 3 \times 1$ | $4 \times 4 \times 1$ | $5 \times 5 \times 1$ | $7 \times 7 \times 1$ |
| PBE | 0.84 | — | 0.80 | 0.81 |
| SCAN | 0.64 | — | 0.63 | 0.54 |
| HSE06 | 1.03 | 0.98 | 0.97 | — |

up to the dense $7 \times 7 \times 1$ grid.

## 4 Plane wave energy cutoff convergence

The plane wave energy cutoff criterion is governed by the ENCUT tag in VASP input files. Figure 2 reports convergence of the total energy of the gN1-G32 model at the full HSE06 level.

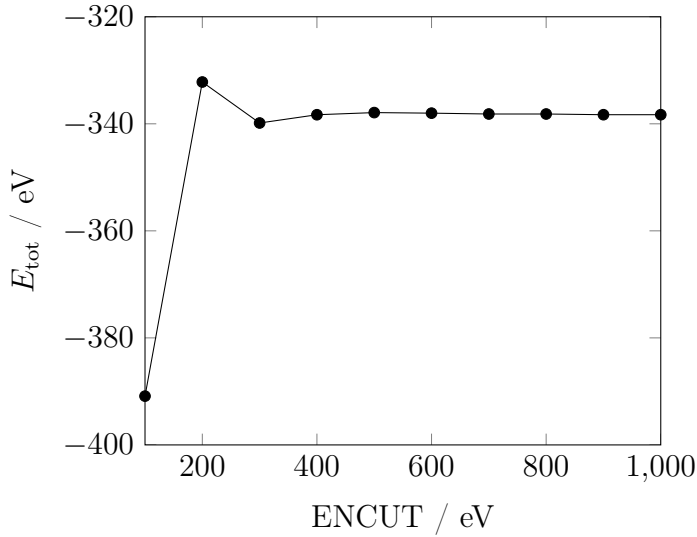

Figure 2: ENCUT convergence for the total energy of adsorbate-free gN1-G32 at the HSE06 level, using single-point calculations on relaxed geometry and a $5 \times 5 \times 1$ $\boldsymbol{k}$ grid.

## 5 Error estimation: exact exchange $\boldsymbol{k}$ point reduction scheme for hybrid DFT geometries

We evaluate the error introduced by the EE $\boldsymbol{k}$ point reduction scheme for hybrid calculations presented in and used throughout the main manuscript. Table 3 compares (purely electronic) adsorption energy values for HSE06 obtained by a) geometry optimization using the $\boldsymbol{k}$ reduction scheme and final single-point energy evaluation on the full grid (denoted HSE06*) and b) both geometry relaxation and final single-point energies on the full grid. We used

4

Table 3: Comparison of total electronic energy values obtained via the HF $k$ point reduction scheme for geometry optimizations (HSE06*) and by full optimization at the HSE06 level (HSE06).

| System | $E_{\text{tot}}^{\text{HSE06*}}$ / eV | $E_{\text{tot}}^{\text{HSE06}}$ / eV | $\Delta E_{\text{tot}}$ / eV |
|---|---|---|---|
| clean NG sheet | -334.8969 | -334.9058 | -0.0089 |
| *O | -342.3237 | -342.3298 | -0.0062 |
| *OH | -347.3367 | -347.3434 | -0.0067 |
| *OOH | -353.4844 | -353.4927 | -0.0083 |

a $4 \times 4 \times 1$ $k$ grid and an energy cutoff of 500 eV for this comparison. Note that the full optimization is extremely computationally demanding, making denser $k$ grids unreasonable to use. The introduced error is *ca.* 0.01 eV, which is noticeable but overall one magnitude lower than that introduced by other factors such as the choice of functional.

## 6    Error estimation: constrained geometry *vs.* full phonon calculation

Tables 4 and 5 list the electronic energy values, free energy corrections and final adsorption free energy values obtained using constrained phonon calculations (backbone constrained, considered only contributions from the adsorbate) and a full phonon calculation (including contributions from the backbone) at the PBE level.    Calculation of the adsorption free

Table 4: Thermodynamic corrections from constrained phonon calculation.

| System | $E_{\text{tot}}$ / eV | $E_{\text{ZPE}}$ / eV | $T \cdot \Delta S$ / eV | $G_{\text{tot}}$ / eV | $G_{\text{ads}}$ / eV |
|---|---|---|---|---|---|
| gN1-G32 | -293.464 | — | — | — | — |
| *O | -298.163 | 0.083 | 0.049 | -298.129 | 2.499 |
| *OH | -303.274 | 0.392 | 0.062 | -302.944 | 0.936 |
| *OOH | -307.427 | 0.466 | 0.149 | -307.110 | 3.933 |

energies used H$_2$O (-13.669 eV) and H$_2$ (-6.506 eV) as reference in both cases. When OP$_{\text{T}}$

Table 5: Thermodynamic corrections from full phonon calculation.

| System | $E_{\text{tot}}$ / eV | $E_{\text{ZPE}}$ / eV | $T \cdot \Delta S$ / eV | $G_{\text{tot}}$ / eV | $G_{\text{ads}}$ / eV |
|---|---|---|---|---|---|
| gN1-G32 | -293.464 | 5.288 | 0.394 | -288.570 | — |
| *O | -298.163 | 5.392 | 0.442 | -293.213 | 2.521 |
| *OH | -303.274 | 5.734 | 0.438 | -297.978 | 1.008 |
| *OOH | -307.427 | 5.811 | 0.509 | -302.125 | 4.024 |

is calculated from these values, the disparity amounts to 0.04 V (0.29 V for the constrained *vs.* 0.33 V for the unconstrained calculation). We therefore suggest that the reported overpotentials are slightly underestimated throughout the manuscript as a result of this approximation.

# 7 Calculation of potential-dependent free energy diagrams

Free energy diagrams were constructed according to the method by Nørskov and co-workers.[3–5] The Standard Hydrogen Electrode (SHE) is defined so that

$$\text{H}_{2(\text{g})} \longleftrightarrow 2\,(\text{H}^+_{(\text{aq})} + \text{e}^-) \qquad (1)$$

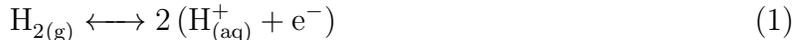

holds true, which corresponds to reaction conditions of pH 0 in solution, $\text{H}_2$ pressure of 1 bar, and room temperature (298.15 K). This allows to define the following reactions (2)–(4) to be in equilibrium at an electrode potential of $U = 0$,

$$2\,\text{H}_2\text{O} + * \longleftrightarrow *\text{OOH} + \frac{3}{2}\text{H}_2 \qquad (2)$$

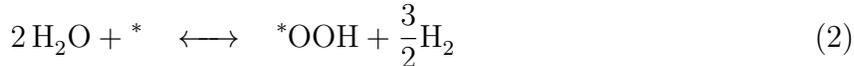

$$\text{H}_2\text{O} + * \longleftrightarrow *\text{OH} + \frac{1}{2}\text{H}_2 \qquad (3)$$

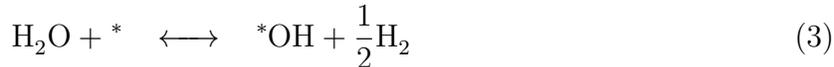

$$\text{H}_2\text{O} + * \longleftrightarrow *\text{O} + \text{H}_2\,, \qquad (4)$$

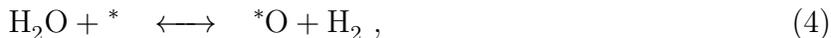

whereby * denotes the lowest-energy adsorption position on the NG backbone. Within this approximation, the references are defined with respect to gas phase $H_2$ and $H_2O$, thus avoiding calculation of the notoriously difficult $^3O_2$ molecule or the radical ˙OH and ˙OOH species in isolation.

Assuming the associative pathway, the individual steps of the catalytic 4e$^-$ ORR pathway on NG can therefore be defined, given in Equations (5) to (8), as

$$O_2 + * + H^+ + e^- \longrightarrow {}^*OOH \tag{5}$$

$$^*OOH + H^+ + e^- \longrightarrow {}^*O + H_2O \tag{6}$$

$$^*O + H^+ + e^- \longrightarrow {}^*OH \tag{7}$$

$$^*OH + H^+ + e^- \longrightarrow H_2O + * \tag{8}$$

Thus, the following equations for the potential-dependent adsorption Gibbs free energies $\Delta G(U)$ of each reaction step[3] can be derived:

$$\Delta G_{OOH}(U) = (G_{OOH} + 2\,G_{H_2O}) - (G_{clean} + 3\,G_{H^+}) - 3\,eU \tag{9}$$

$$\Delta G_{O}(U) = (G_{O} + G_{H_2O}) - (G_{clean} + 2\,G_{H^+}) - 2\,eU \tag{10}$$

$$\Delta G_{OH}(U) = (G_{OH} + G_{H_2O}) - (G_{clean} + G_{H^+}) - eU \tag{11}$$

$$\tag{12}$$

Gibbs free energy results are calculated as $G_x(U_0) = E_x + E_{ZPE} + T \cdot S_x$, with the total SCF energy from DFT calculations $E_x$, zero-point energy $E_{ZPE}$ and the entropic correction $T \cdot S_x$ at room temperature $T = 298.15\,K$ and pH 0 calculated from harmonic vibrational frequency analysis.

# 8 Free energy diagrams

Free energy diagrams at 0 V and at the onset and at the ORR onset potential are constructed for the PBE, BEEF-vdW, TPSS, SCAN, PBE0, HSE06, and B3LYP functionals, see Figure 3.

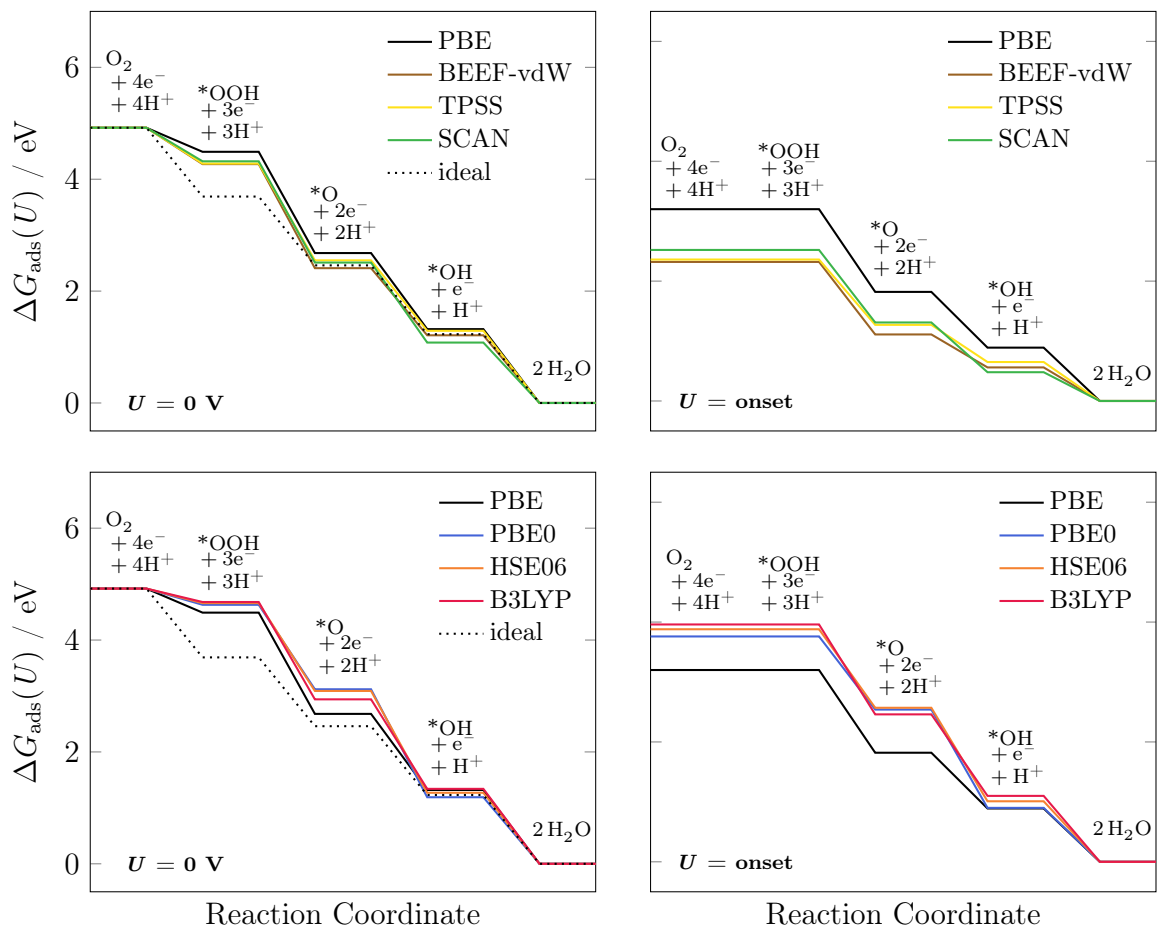

Figure 3: Free energy diagrams for gN1-G32 at 0 V (left) and at the onset electrode potential (right) obtained using GGA and meta-GGA functionals (top) as well as PBE and hybrid functionals (bottom). The ideal ORR catalyst with a free energy change of 1.23 V at each reaction step is shown as a dashed line.

# 9 Influence of dispersion correction

Table 6: Calculated total energy difference $\Delta E = E_{\text{tot}}^{\text{top}} - E_{\text{tot}}^{\text{bridge}}$ of graphene sheets with top- and bridge-bound adatoms, using various DFT functionals, SIC, and DMC. For $M$ point results, the relative deviation with respect to DMC is given in brackets. This table contains additional values for dispersion-corrected functionals not given in the main manuscript. Dispersion correction is found to play an insignificant role.

|            | $\Delta E_{\text{D}}$ / eV | | | | |
|------------|---------|-------------|--------------|----------------------|-----------|
| Method     | Hsing[6,a] | $M$ point[b] | full $k$ grid[c] | opt. + full $k$ grid[d] | Janesko[7,e] |
| DMC        | 1.360   | –           | –            | –                    | –         |
| SIC        | –       | 1.47 (+8 %) | –            | –                    | –         |
| PBE0       | –       | 1.29 (-5 %) | 1.10         | 1.09                 | –         |
| HSE06      | –       | 1.28 (-6 %) | 1.10         | 1.10                 | 1.31      |
| HSE06-D3   | –       | 1.26 (-7 %) | 1.26         | 1.05                 | –         |
| B3LYP      | –       | 0.99 (-27 %)| 0.80         | 0.80                 | 1.10      |
| TPSS       | –       | 0.97 (-29 %)| 0.78         | 0.79                 | 0.93      |
| SCAN-rVV10 | –       | 0.941 (-31 %)| 0.959       | 0.923                | –         |
| SCAN       | –       | 0.92 (-32 %)| 0.96         | 0.92                 | –         |
| PBE        | 0.74    | 0.92 (-32 %)| 0.75         | 0.75                 | 0.91      |
| PBE-D3     | –       | 0.91 (-33 %)| 0.72         | 0.72                 | –         |
| BEEF-vdW   | –       | 0.82 (-40 %)| 0.65         | $0.63 \pm 0.20$[f]   | –         |

[a] $M$ point evaluated for geometries provided by Hsing et al.[6]
[b] $5 \times 5 \times 1$ $k$ grid evaluated for geometries provided by Hsing et al.[6]
[c] $5 \times 5 \times 1$ $k$ grid evaluated for geometries optimized using the respective density functional.
[d] Value obtained from $M(0.5, 0)$ point DMC calculation.
[e] Standard deviation based on an ensemble size of 2000.

# 10 Reda *et al.* solvation contribution reproduction with HSE06

To complete the thermodynamic description of the models presented in this study, the influence of solvation on the stability of the ORR intermediates is considered. To this end, the influence of a single layer of 8 $H_2O$ molecules (corresponds to $\theta = \frac{2}{3}$ monolayer) on the binding energy of the adsorbate is assessed. A single layer of water was recently found to give converged solvation energy results in a study by Reda *et al.* on a similar NG model system.[8] The solvation stabilizations obtained from the HSE06 functional in these calculations are summarized in Table 7. Figure 4 compares the resulting free energy diagrams to

Table 7: Solvation stabilization by 8 explicit $H_2O$ molecules for ORR intermediates on NG. Negative values indicate stabilization.

|  Intermediate | $\Delta E_{\text{solv}}$ / eV gN1-G32p |
| --- | --- |
| *O | -0.38 |
| *OH | -0.12 |
| *OOH | -0.14 |

gas phase results from gN1-G32p. Inclusion of solvation stabilization in the free energy description presented above will lead to a reduction of the thermochemical overpotentials from 0.97 to 0.82 V. This is exclusively due to stabilization of the *OOH intermediate whose free energy governs the onset potential. Note that the *O intermediate experiences significant stabilization as well which could be significant for systems where the *O intermediate is the bottleneck step.

# 11 Illustrations of flake models

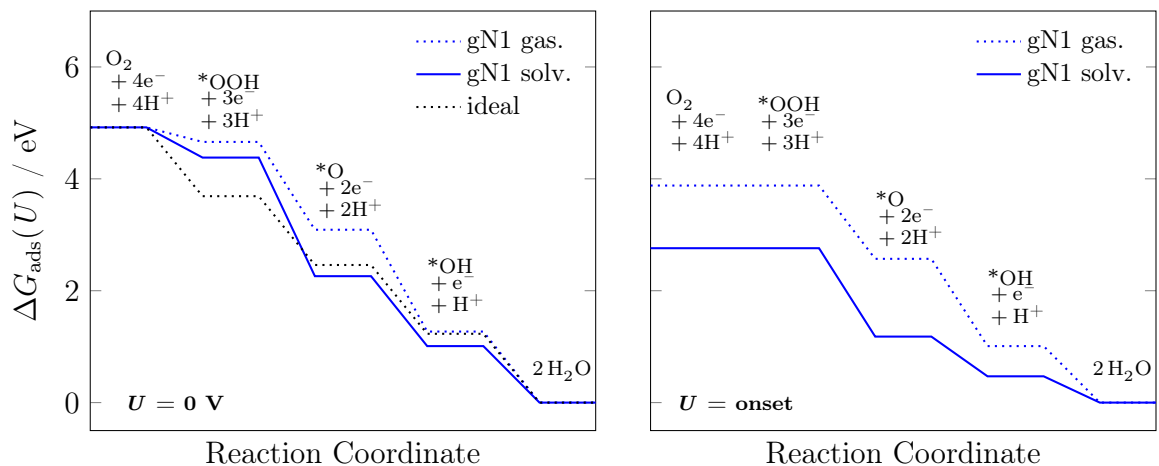

Figure 4: Free energy diagrams for gN1-G32p blue at 0V (left) and the onset electrode potential (right) obtained with HSE06. Gas.: pure gas phase values; solv.: including solvation correction from 8 explicit $H_2O$ molecules.

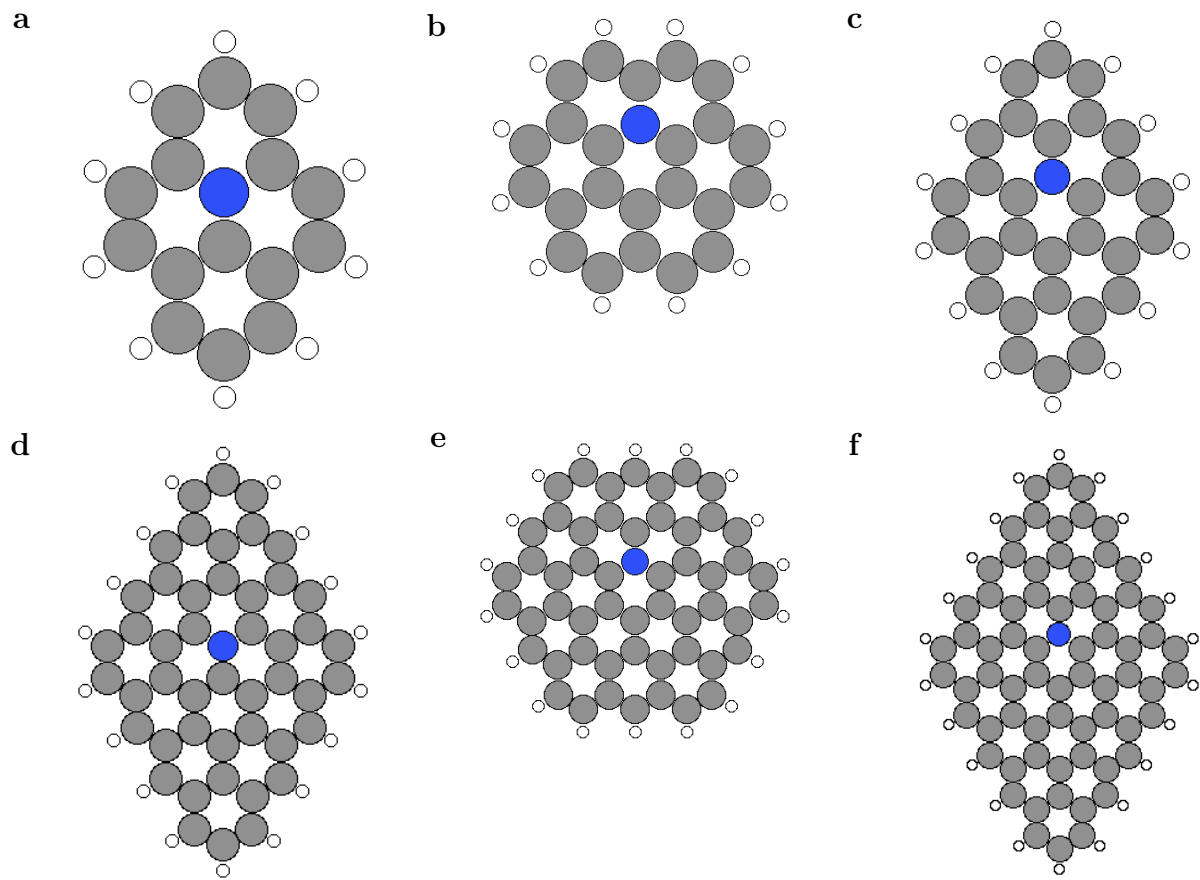

Figure 5: Illustration of unconstrained flake models with **a** 16, **b** 24, **c** 30, **d** 48, **e** 54, and **f** 70 C and N atoms.

11

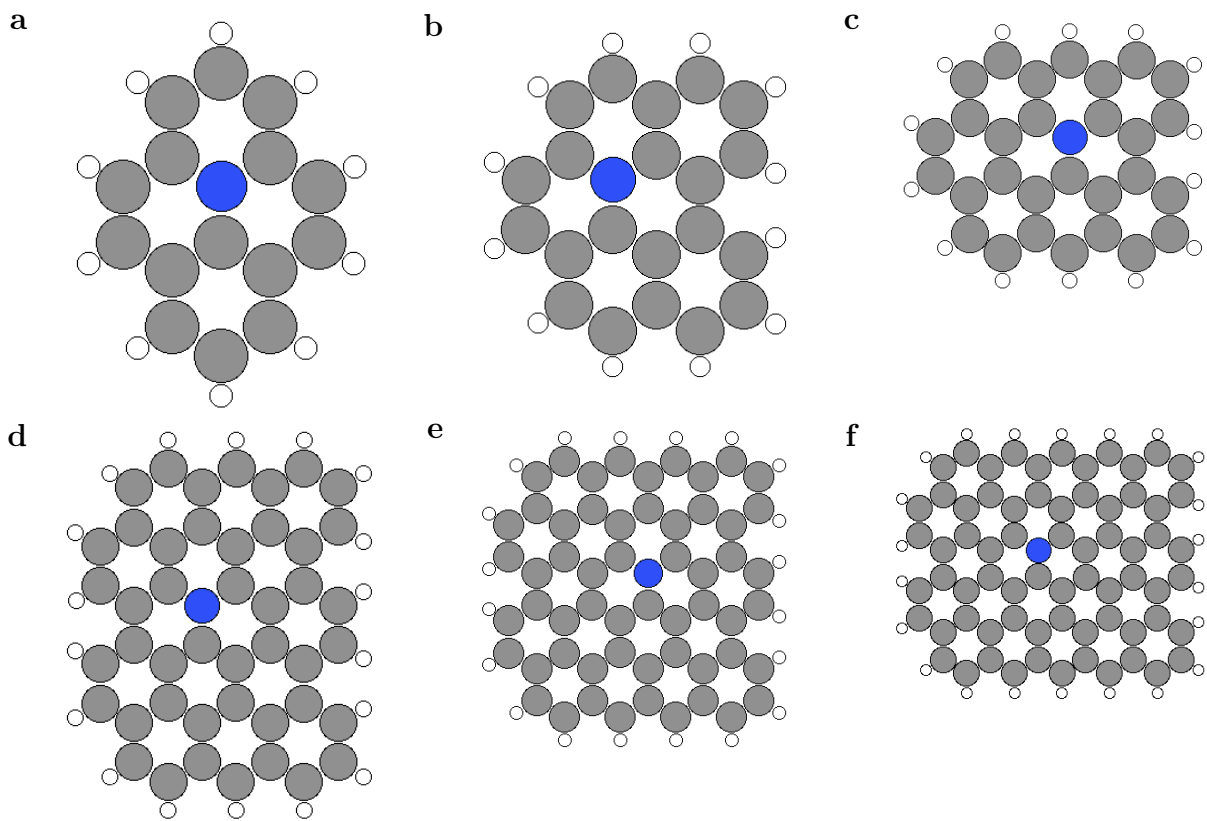

Figure 6: Illustration of flake models generated from periodic structures with **a** 16, **b** 24, **c** 32, **d** 48, **e** 60, and **f** 72 C and N atoms.

## 12   Free energy diagrams for flake models

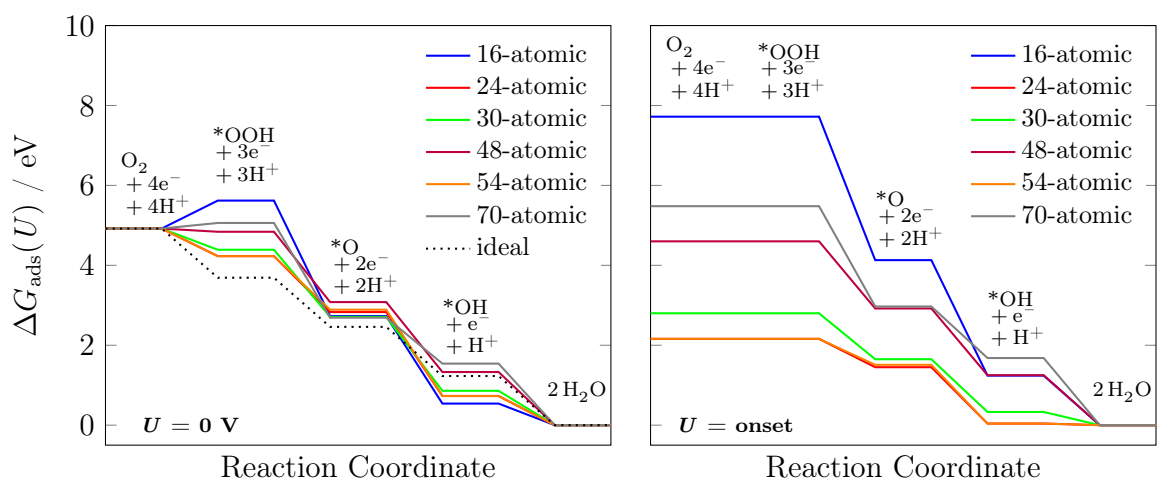

Figure 7: Free energy diagrams for round and diamond shaped flakes that are allowed to fully relax during the calculations.

12

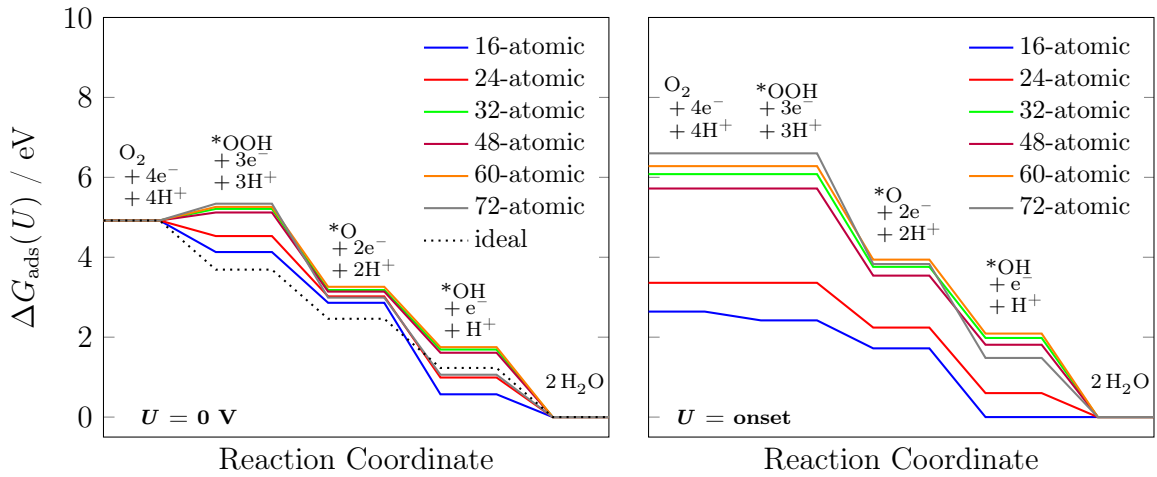

Figure 8: Free energy diagrams for flakes generated from periodic models and capped with hydrogen atoms. Only the hydrogen atoms are allowed to relax during the calculation while the C/N backbone is kept rigid in the atomic configuration of the periodic model system.

13



\end{document}